\documentclass[fleqn,usenatbib]{mnras}

\usepackage{newtxtext,newtxmath}

\usepackage[T1]{fontenc}

\DeclareRobustCommand{\VAN}[3]{#2}
\let\VANthebibliography\thebibliography
\def\thebibliography{\DeclareRobustCommand{\VAN}[3]{##3}\VANthebibliography}

\usepackage{graphicx}	
\usepackage{amsmath}	
\usepackage{bm}
\usepackage{url}

\title[Effective dust growth in laminar CPDs]{Effective dust growth in laminar circumplanetary discs with magnetic wind-driven accretion}

\author[Y. Shibaike and S. Mori]{
Yuhito Shibaike$^{1}$\thanks{E-mail: yuhito.shibaike@unibe.ch (YS)}
and Shoji Mori$^{2}$
\\
$^{1}$Physikalisches Institut \& NCCR PlanetS, Universitaet Bern, CH-3012 Bern, Switzerland\\
$^{2}$Astronomical Institute, Tohoku University, 6-3 Aramaki, Aoba-ku, Sendai 980-8578, Japan\\
}

\date{Accepted XXX. Received YYY; in original form ZZZ}

\pubyear{2022}

\begin{document}
\label{firstpage}
\pagerange{\pageref{firstpage}--\pageref{lastpage}}
\maketitle

\begin{abstract}
It has been considered that large satellites around gas planets form in-situ circumplanetary discs (CPDs). However, dust particles supplied into CPDs drift toward the central planets before they grow into satellitesimals, building blocks of the satellites. We investigate the dust growth in laminar CPDs with magnetic wind-driven accretion. In such laminar discs, dust particles can settle onto the mid-plane and grow large by mutual collision more efficient than in classical turbulent CPDs. First, we carry out 3D local MHD simulations of a CPD including all the nonideal MHD effects (Ohmic resistivity, Hall effect and ambipolar diffusion). We investigate if the disc accretion can be governed by magnetic wind-driven accretion and how laminar the disc can be, in a situation where the magnetic disc wind can be launched from the disc. Second, we model 1D steady CPDs consistent with the results of the MHD simulations and calculate the steady radial distributions of the dust profiles in the modelled discs, taking account of the collisional growth, radial drift, fragmentation, and vertical stirring by the Kelvin-Helmholtz instability. We show that satellitesimals can form in such CPDs if the dust-to-gas mass ratio of the inflow to the discs is larger than 0.02, which is 50 times smaller than the critical value in turbulent CPDs. This condition can be satisfied when enough amount of dust piles up at the gas pressure bump created by the planets. This result shows that satellitesimals would form in laminar CPDs with magnetic wind-driven accretion.
\end{abstract}




\section{Introduction}
\label{introduction}
It is generally accepted that the large satellites around Jupiter and Saturn formed from small solid materials inside their circumplanetary discs (CPDs), byproducts of the gas accretion of the gas planets, like planets form in protoplanetary discs (PPDs). There are mainly two in-situ formation scenarios for the major satellites: the satellitesimal-accretion scenario \citep[e.g.,][]{mos03a,can06} and the pebble-accretion scenario \citep{shi19,ron20}. The former needs a lot of satellitesimals (km sized or larger), building blocks of the satellites, but the satellitesimal formation is not easy. Dust particles supplied to the disc quickly grow to pebbles (cm-m sized) by their mutual collision, but they then fall into the central planets due to the headwind receiving from the gas rotating with sub-Keplerian speed. \citet{shi17} found that the in-situ formation of satellitesimals is difficult unless the dust-to-gas mass ratio of the inflow onto the CPDs is larger than unity, which is not realistic \citep{hom20,szu21}. On the other hand, \citet{dra18b} and \citet{bat20} argued that if the discs have strong radial outflows on the mid-planes, the particles can pile up at the points where the inward drift and the outward advection are balanced and satellitesimals form by the streaming instability from the gathered particles. In this work, we propose another solution to avoid the drift barrier to the satellitesimal formation.

Previous works modelled their CPDs as the classical $\alpha$ disc models \citep{sha73}, where their dynamics and evolution are governed by single parameter $\alpha$, the strength of the turbulence of the discs, in the literature of the satellite formation \citep[e.g.,][]{shi17}. However, the $\alpha$ parameter can be distinguished into two different kinds of parameters based on the physical properties: the efficiency of the angular momentum transport, $\alpha_{\rm acc}$, and the diffusion strength of dust particles, $\alpha_{\rm diff}$. The former determines the evolution of the gas accretion disc and the latter excites the motion of dust (and gas) particles \citep{you07}. If $\alpha_{\rm acc}$ is high, the gas surface density is low and thereby the drift time scale of pebble can be long. On the other hand, if $\alpha_{\rm diff}$ is low, the collision rate remains high as the particles settle on the mid-plane, provided the collision speed is not determined by the diffusion, and thus particles grow quickly. \citet{dra18a} found that this is effective for the planetesimal formation in PPDs. In this paper, we investigate if it also works in the satellitesimal formation.

In the case of PPDs, the two values of $\alpha$ can be different when the disc is governed by the layered accretion or the magnetic wind-driven accretion. It has been considered that the most effective accretion mechanism of PPDs is the magneto-rotational instability (MRI) \citep{bal91,bal98}. However, \citet{gam96} found that one of the nonideal magneto-hydrodynamic (MHD) effects, Ohmic resistivity prevents the MRI around the mid-plane, which is called ``dead-zone'', while it is active only at the surface of the disc. In that case, the disc becomes a layered structure and the gas flows inward only at the upper layer. \citet{oku11} carried out 3D ohmic-resistive MHD simulations and showed that the layered accretion occurs and $\alpha_{\rm acc}\sim5-30~\alpha_{\rm diff}$ in the inner regions of PPDs. On the other hand, \citet{bai15} calculated 3D MHD simulations including all three nonideal MHD effects (Ohmic resistivity, Hall effect and ambipolar diffusion) and found that the MRI is dead not only on the mid-plane but also at the whole height of the discs\footnote{\citet{bai13} also found that the MRI is dead at the whole height of PPDs by MHD simulations including Ohmic resistivity and ambipolar diffusion.}. Therefore, PPDs would be laminar rather than turbulent. In that case, the angular momentum is transported at the surface of the disc where the large-scale magnetic field lines are bent. This accretion mechanism is called magnetic wind-driven accretion, because gas escapes from the disc along with the magnetic field lines, which is called disc wind\footnote{The word ``disc wind'' is used for general mass loss from PPDs, and it launches not only from magnetic wind-driven accretion discs but also from MRI turbulent discs \citep{suz16}.} (see also Section~\ref{MHDresults} for detailed explanations). \citet{bai15} argued that $\alpha_{\rm acc}\gtrsim10~\alpha_{\rm diff}$ by a simple estimate from the outcome value of the rms vertical velocity fluctuation of the simulation results.

Most of the numerical simulations of CPDs calculated in previous works are not MHD but hydrodynamics (HD). \citet{mac06} and \citet{gre13} carried out global 3D MHD simulations of CPDs, though the former was ideal, and the latter included only Ohmic resistivity. They found that jet structures can form around the centre of the CPDs like the star formation process. In the simulation by \citet{gre13}, MRI is active at the surface layers of the CPDs but is dead inside the discs. It has also been considered that MRI is dead in most of the regions of CPDs other than the surfaces because of the short typical length scale of the discs \citep{fuj14,kei14,tur14,fuj17}. These works calculated the Elsasser number and the plasma beta of CPDs in 1D disc models. In this work, we carry out 3D local nonideal MHD simulations of CPDs to investigate if the discs are laminar and the magnetic wind-driven accretion dominates the disc accretion. This is the first work that carries out 3D MHD simulations of CPDs with all the nonideal MHD effects, but the simulations are performed not in global but in local shearing boxes.

In Section~\ref{MHD}, we perform 3D local nonideal MHD simulations of a CPD. We consider situations where the magnetic disc wind can be launched from the disc and investigate if the disc state can be governed by the magnetic wind-driven accretion and $\alpha_{\rm acc}$ is much larger than $\alpha_{\rm diff}$. In Section~\ref{formation}, we calculate the evolution of dust particles in 1D CPDs modelled to be consistent with the results of the MHD simulations and investigate the condition for the satellitesimal formation. In Section~\ref{discussion}, we estimate the region where the disc wind is launched against the vertical gas inflow ignored in the MHD simulations. We then discuss the results of MHD simulations comparing with those of the previous works. We also discuss the feasibility of the obtained condition for the satellitesimal formation; it can be reached even when the gas gap is open. We also perform the parameter studies of the satellitesimal formation. We finally conclude this work in Section~\ref{summary}.

\section{Nonideal MHD simulations of CPDs} \label{MHD}
\subsection{Setups} \label{simsetup}
We perform nonideal MHD simulations of a CPD with an open source MHD code \texttt{Athena}\footnote{\url{https://princetonuniversity.github.io/Athena-Cversion/}} \citep{sto08}. The simulations are performed in local 3D stratified shearing boxes, reproducing four different locations in a CPD, $r=3$, $10$, $30$ and $100~R_{\rm J}$, where $r$ is the distance from the central planet and $R_{\rm J}$ is the Jupiter radius. 

Here we put two simplifications in the CPD simulations. First, we do not set the vertical gas inflow to CPDs as the boundary condition. The inflowing boundary condition in such shearing box simulations has not been investigated well so far. In addition, the inflowing boundary condition can easily break the simulation. Such disc-wind simulations taking into account the inflow effects is a future challenge. Nevertheless, we discuss if the disc has an enough potential to launch the wind even with the inflow by using the simulation results, in Section~\ref{region}. Second, we do not consider any horizontal gas flow either, because shearing box simulations do not have the concept of inward/outward directions.

The simulation boxes are orbiting the central planet with Kepler angular velocity $\Omega_{\rm K}=\sqrt{GM_{\rm p}/r^{3}}$, where $G$ is the gravitational constant and $M_{\rm p}=1M_{\rm J}$ (one Jupiter mass) is the mass of the central planet. The simulations are carried out with Cartesian coordinates $(x, y, z)$ for the radial, azimuthal and vertical dimensions. Previous MHD simulations in the inner regions of PPDs showed that the final properties do not depend on the horizontal box sizes and resolutions \citep{bai13}, and we have confirmed that it is also true in our cases. Thus, we use simulation boxes with the vertical and horizontal box sizes of 16 $H_{\rm g}$ and 0.5 $H_{\rm g}$ and the resolutions of 25 and 8 cells per $H_{\rm g}$, respectively, where  $H_{\rm g}$ is the gas scale height of the CPDs.  The gas scale height is $H_{\rm g}=c_{\rm s}/\Omega_{\rm K}$, were $c_{\rm s}$ is the sound speed, and $c_{\rm s}=\sqrt{k_{\rm B}T/m_{\rm g}}$, where $k_{\rm B}$, $T$ and $m_{\rm g}=3.9\times10^{-24}~{\rm g}$ are the Boltzmann constant, disc temperature and mean molecular mass, respectively.

We set the gas temperature profile of the CPD as $T=160(r/10 R_{\rm J})^{-3/4}~{\rm K}$, which is consistent with the ice mass fractions of the Galilean satellites \citep[e.g.,][]{kus05}, and assume the gas is isothermal in each simulation box. Although the assumed temperature is lower than that of turbulent discs, we note that the heating can occur at high altitude in magnetic wind-driven accretion discs, and the mid-plane temperature can be lower \citep{mor19a}. As the initial conditions, we set the gas surface density as $\Sigma_{\rm g}=1000(r/10R_{\rm J})^{-3/4}~{\rm g~cm}^{-2}$ (see Section \ref{comparison} for the reason why we choose this profile) and the density distribution in hydrostatic equilibrium. We fix the dust-to-gas ratio of $0.1~{\rm\mu{\rm m}}$-sized dust particles as $f_{\rm dg}=1\times10^{-6}$.

We also set the initial magnetic field as ${\bm B}_{0}=(0, 0, B_{0})$, where $B_{0}$ is characterized by a parameter $\beta_{0}=8\pi\rho_{\rm g,0}c_{\rm s}^{2}/B_{0}^{2}$, which is assumed as $\beta_{0}=10^{5}$ in this work. Depending on the direction of $B_{\rm 0}$, the Hall effect amplifies or damps the magnetic field. We here take $B_{\rm 0}$ to be aligned with the disc rotation vector, in which the magnetic field is amplified by the Hall effect.

We give the magnetic diffusivities of the nonideal MHD effects with calculating the balance between the ionization and recombination, basically following \citet{mor19a}. The magnetic diffusivities depend on the number densities of charged particles (electrons, ions and charged dust particles). The ionization sources are cosmic rays, stellar X-rays and radionuclides. The cosmic-ray ionization rate profile is described in \citet{san00} which is based on \citet{ume81}. The ionization rate of the X-ray follows the fitting formula consistent with \citet{ige99} and \citet{bai09} but we neglect components that directly reach the disc surface. We set the X-ray temperature to be $5~{\rm keV}$ and the X-ray luminosity to be $2\times10^{30} {\rm erg~s^{-1}}$. Also, the location of the CPD is assumed to be $5.2~ {\rm au}$ from the protosun. The latter is the median value of the luminosity of the Solar-mass stars in the Orion Nebula Cluster \citep{gar00}. The radionuclide ionization rate is set as a constant value of $7.6 \times 10^{-19} {\rm s^{-1}}$ \citep{ume09}. We consider the recombination in gas-phase and on dust particles, where we represent ions and ionized dust particles by a single species. We limit the total diffusivities to 300 $c_{\rm s}H_{\rm g}$ to avoid the calculation time step to be too short.

We also consider the ionization by far-ultraviolet radiation (FUV) above its penetration depth. The stellar FUV rays are scattered around the wall of the gaps and partially enters the CPDs, which is reduced to 4\% of the stellar luminosity \citep{tur12}. We assume the stellar FUV luminosity as 2$\times10^{31}$ erg s$^{-1}$, and so the FUV flux reaches to the CPDs is equivalent to the luminosity of 8$\times10^{29}$ erg s$^{-1}$. The stellar FUV luminosity is set as a higher value than that of the current sun because it should have been higher in the past \citep[e.g.,][]{yan12}. The penetration depth is then about 0.03 g cm$^{-2}$, following the ionization models in \citet{per11b}. We obtain the ionization fraction of $3\times 10^{-5}$ in the FUV layer with a similar treatment used in \citet{bai13}.

\subsection{Results} \label{MHDresults}
\subsubsection{Magnetic disc wind of CPDs} \label{wind-cpd}
\begin{figure*}
\centering
\includegraphics[scale=0.4]{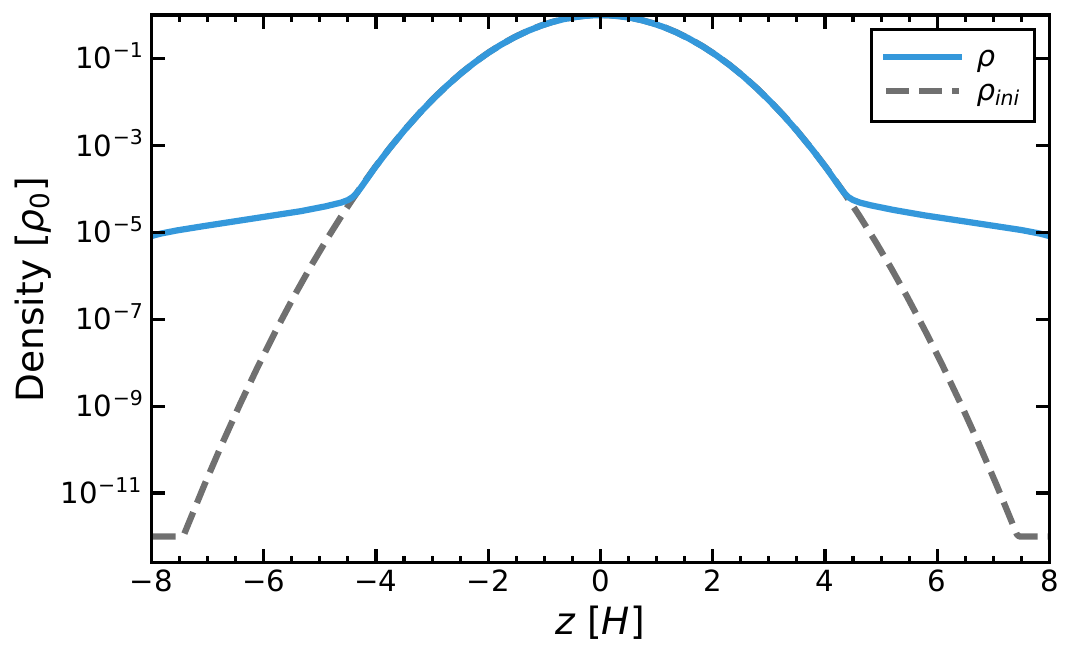}
\includegraphics[scale=0.4]{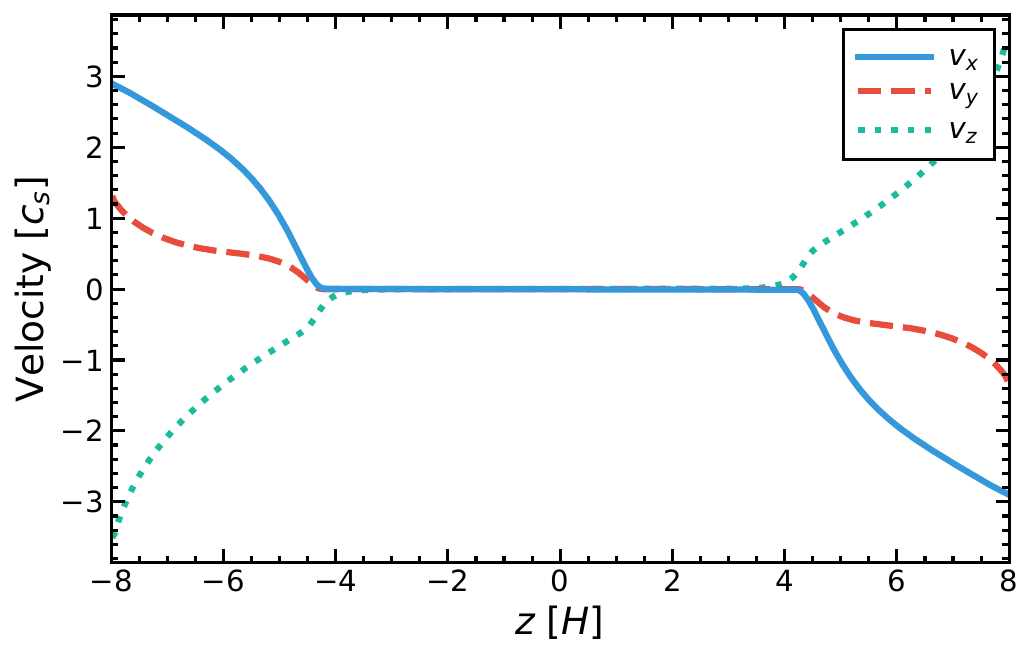}\\
\includegraphics[scale=0.4]{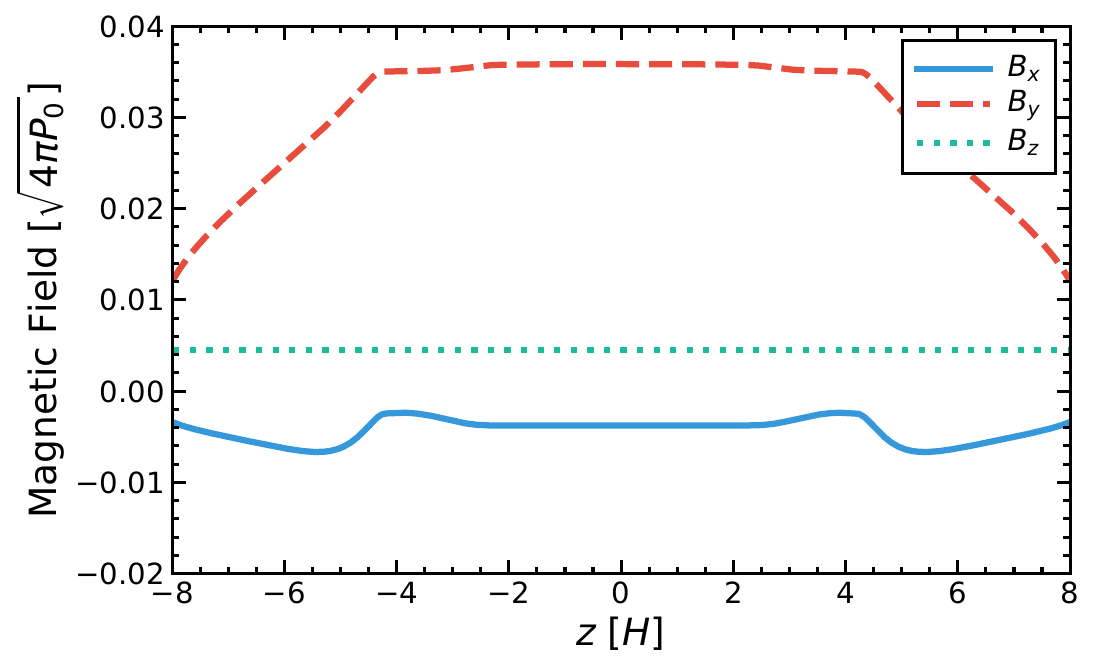}
\includegraphics[scale=0.4]{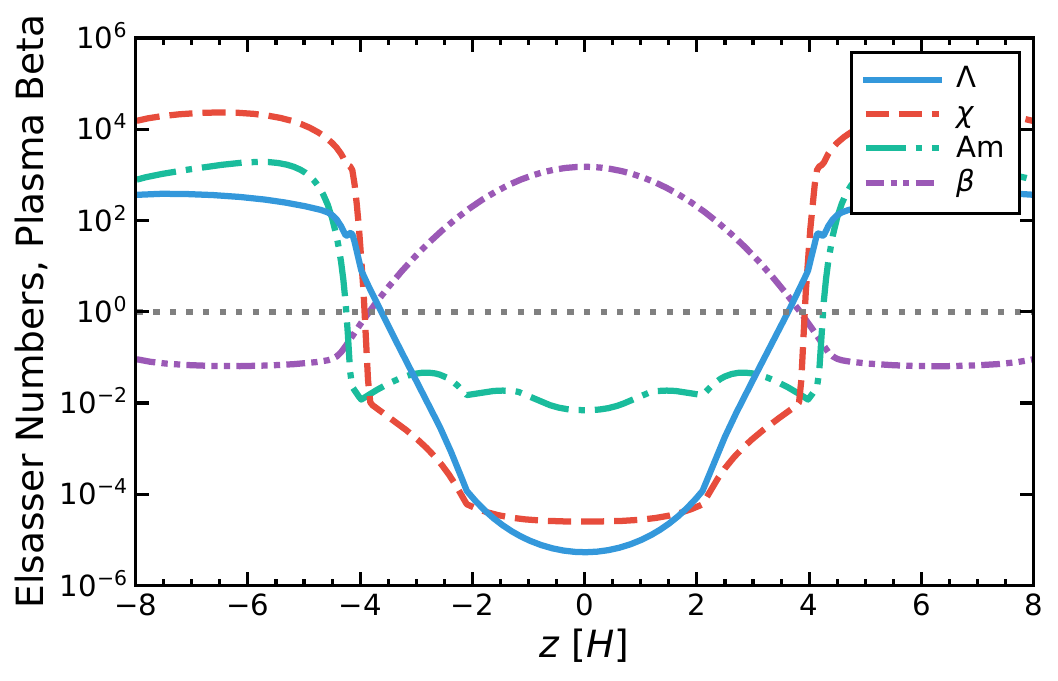}
\caption{Vertical profiles of the temporal and horizontal average for a representative run at $r$=10$R_J$. Upper left: gas density at the final state ($\rho$, solid) and initial state ($\rho_{\rm ini}$, dashed), normalized by the initially set density on the mid-plane, $\rho_0$. Upper right: three components of the gas velocity ($v_x$, $v_y$, $v_z$). Lower left: three components of the magnetic field. Lower right: Elsasser numbers for Ohmic resistivity ($\Lambda$), Hall effect ($\chi$) and ambipolar diffusion ($\rm Am$) and plasma beta ($\beta$).}
\label{fig:sim-prof}
\end{figure*}

The obtained results of the four simulations at the different places are qualitatively similar, so we focus on the simulation results at $r=10 R_{\rm J}$. Figure~\ref{fig:sim-prof} represents the vertical profiles of the density, velocity and magnetic fields of the simulation result. These are the temporal and horizontal average over the last 10 orbits of the total 50 orbits. However, after the vertical profiles reach these states shown in the figure, the profiles are almost steady and have only small variation. These vertical profiles are like those of PPDs with wind-driven accretion \citep[e.g.,][]{bai13}.

The upper two panels of Fig.~\ref{fig:sim-prof} show that the magnetic disc wind is launched from the surfaces of the CPD ($|z|=4 H_{\rm g}$). At $|z|>4 H_{\rm g}$, the density profile deviates from the equilibrium profile given as the initial profile (upper left panel) and the gas flows out vertically (upper right). This is explained by the behavior of the magnetic field (lower left). The gas is pushed up to the surfaces of the CPD  by the toroidal magnetic field amplified at $|z|<4 H_{\rm g}$. The gas then gains angular momentum via the poloidal field threading the disc around its surfaces, thereby flowing away as the disc wind. The disc wind is launched continuously in this situation.

The magnetic field profile can be explained by the Elsasser numbers of the diffusivities (lower right). The Elsasser number of a diffusivity $\eta$ is defined as 
\begin{equation}
\Lambda=\frac{v_{\rm A}^2}{\Omega_{\rm K}\eta},
\end{equation}
where $v_{\rm A}$ is the Alfv\'{e}n speed. When the Elsasser number is less than unity, the nonideal MHD effects are stronger than the magnetic induction due to the shear flow. Around the mid-plane, the Ohmic diffusion is the dominant MHD process. Perturbations of magnetic field are quickly diffused, and so the magnetic field is constant. In addition, ambipolar diffusion dominates over the magnetic induction within $4H_{\rm g}$. For these reasons, the MRI turbulence is fully suppressed below $4H_{\rm g}$, and therefore the CPD is laminar. Nevertheless, the Hall effect at $z=2$--$4 H_{\rm g}$ amplifies the magnetic field with the shear flow \citep[i.e., Hall-shear instability;][]{kun08}. The amplified toroidal field diffuses into the CPD, so that the magnetic field is still strong even in the disc.

\subsubsection{Magnetic wind-driven accretion} \label{wind-driven}
The disc wind flowing out from the surfaces of CPDs takes the disc angular momentum from the gas around the surfaces. The gas losing their angular momentum accretes toward the central planets as the reaction of the disc wind. This mechanism efficiently drives disc accretion even in the presence of the dead zones. We estimate the accretion rates by measuring the imposed stress. The total gas accretion rate, $\dot{M}_{\rm g}$, is the sum of the rates due to the radial and vertical angular momentum transfer, $\dot{M}_{\rm acc,r}$ and $\dot{M}_{\rm acc,z}$ \citep{bai13,mor19a}\footnote{The gas accretion rate driven by the vertical angular momentum transfer, $\dot{M}_{\rm acc,z}$, is not the mass escaping rate from the CPDs as disc wind.},
\begin{equation}
\begin{split}
\dot{M}_{\rm g}&=\dot{M}_{\rm acc,r}+\dot{M}_{\rm acc,z} \\
&=\dfrac{2\pi}{\Omega_{\rm K}}\int_{-z_{\rm b}}^{z_{\rm b}} T_{xy}dz~+~\dfrac{8\pi}{\Omega_{\rm K}}r|T_{zy}|_{z=z_{\rm b}},
\label{mdotg}
\end{split}
\end{equation}
where $T_{xy}=\rho_{\rm g}v_{x}v_{y}-B_{x}B_{y}/(4\pi)$ and $T_{zy}=-B_{y}B_{z}/(4\pi)$ are the Maxwell stresses in the radial and vertical direction, respectively. Here, we assume the height of the base of the wind as $z_{b}=4H_{\rm g}$. The gas accretion rates in the four simulations are about $\dot{M}_{\rm g}=0.1$--$1~M_{\rm J}~{\rm Myr}^{-1}$ (Table~\ref{tab:MHD}). We also find that $\dot{M}_{\rm g}$ is larger as the distance from the planet is further. This is because the outer region of the disc is relatively well ionized.

We then define the efficiency of the angular momentum transport as
\begin{equation}
\label{alpha-acc}
\alpha_{\rm acc}\equiv\frac{\dot{M}_{\rm g}}{3\pi\Sigma_{\rm g}c_{\rm s}H_{\rm g}},
\end{equation}
to be consistent with the steady accretion disc model in Section~\ref{formation} (see equation~(\ref{sigmag})). We then find that $\alpha_{\rm acc}=2\times10^{-3}$--$2\times10^{-2}$ at $r=3$--$100~R_{\rm J}$ (Table~\ref{tab:MHD}). We also find that $\alpha_{\rm acc}$ is larger as the distance from the planet is further as well as $\dot{M}_{\rm g}$.

Besides, $\alpha_{\rm diff}$ determines the dust vertical distribution and the relative velocity among the particles (see equations~(\ref{Hddiff}) and (\ref{vt}) below), which is assumed to be equal to $\alpha_{\rm acc}$ in the conventional disc models. We measure the diffusion strength as $\alpha_{\rm diff}=\langle\delta v_{{\rm g},z}^{2}\rangle/c_{\rm s}^{2}$, where $\langle\delta v_{{\rm g},z}^{2}\rangle$ is the squared vertical velocity dispersion averaged spatially (see Table~\ref{tab:MHD}). We take the volume average within $z=\pm 2 H_{\rm g}$. The gas motion causes dust motion, so that $\alpha_{\rm diff}$ determines the dust scale height and affects the collision speed among the dust particles. In our simulations, the gas flows near the mid-plane are quiescent and so $\alpha_{\rm diff}$ is significantly low compared to $\alpha_{\rm acc}$. The ratio between the two alpha values is $\alpha_{\rm diff}/\alpha_{\rm acc}=10^{-5}$--$10^{-4}$.

\begin{table*}
\centering
\caption{Input parameters and results of the nonideal MHD simulations.}
\label{tab:MHD}
\begin{tabular}{cccccc}
\hline
Symbol & \multicolumn{4}{c}{Value} & Unit \\
\hline\hline
$r$ & 3 & 10 & 30 & 100 & $R_{\rm J}$ \\
$\Sigma_{\rm g}$ & 2467 & 1000 & 438.7 & 177.8 & ${\rm g~cm^{-2}}$ \\
$T$ & 395 & 160 & 70.2 & 28.5 & ${\rm K}$ \\
\hline
$\dot{M}_{\rm acc,r}$ & 0.0090 & 0.012 & 0.22 & 0.76 & $M_{\rm J}~{\rm Myr}^{-1}$ \\
$\dot{M}_{\rm acc,z}$ & 0.11 & 0.14 & 0.22 & 0.29 & $M_{\rm J}~{\rm Myr}^{-1}$ \\
$\dot{M}_{\rm g}$ & 0.12 & 0.15 & 0.44 & 1.1 & $M_{\rm J}~{\rm Myr}^{-1}$ \\
$\alpha_{\rm acc}$ & $2.5\times10^{-3}$ & $3.1\times10^{-3}$ & $9.1\times10^{-3}$ & $2.2\times10^{-2}$ & - \\
$\alpha_{\rm diff}$ & $7.8\times10^{-8}$ & $7.4\times10^{-8}$ & $8.8\times10^{-7}$ & $4.8\times10^{-7}$ & - \\
$P_{\rm mag}/P_{\rm mid}$ & $3.0\times10^{-4}$ & $6.1\times10^{-4}$ & $2.1\times10^{-3}$ & $5.0\times10^{-3}$ & - \\
\hline
\end{tabular}
\end{table*}

\section{Satellitesimal formation} \label{formation}
\subsection{Methods} \label{methods}
We calculate the growth and drift of dust particles in 1D gas disc models of CPDs. We treat the discs and the distribution of the particles as steady states in this section because the time scale of the dust growth and radial motion is much shorter than that of the evolution of CPDs. The evolution time scale of CPDs should have been determined by that of gas inflow, in other words, that of the parental protoplanetary gas discs.

\subsubsection{Gas disc model} \label{discmodel}
First, we model the structure of CPDs by 1D steady disc model to be consistent with the results of the nonideal MHD simulations. The results of our MHD simulations suggest that the gas from the PPDs falls onto the outside of $r_{\rm out}=30~R_{\rm J}$ (see Section~\ref{region}). We assume that the gas coming from the PPD falls onto the outer region of the CPD ($r>r_{\rm out}$) and all of the infalling mass steadily inflows into the inner region. Therefore, at least in the inner region of the CPD that we focus on ($r\le r_{\rm out}$), the gas accretes inward. We then model the gas disc as
\begin{equation}
\Sigma_{\rm g}=\dfrac{\dot{M}_{\rm g}\Omega_{\rm K}}{3\pi\alpha_{\rm acc}c_{\rm s}^{2}}.
\label{sigmag}
\end{equation}
We set $\dot{M}_{\rm g}=0.15~M_{\rm J}~{\rm Myr}^{-1}$, $\alpha_{\rm acc}=3.1\times10^{-3}$, and $\alpha_{\rm diff}=7.4\times10^{-8}$, which are the values at $r=10~R_{\rm J}$ in our MHD simulations (Table~\ref{tab:MHD}), and we assume that they are uniform in the disc. We also assume that $T=160(r/10R_{\rm J})^{-3/4}~{\rm K}$, which is equal to the setting of the MHD simulations. In Sections \ref{non-uniform} and \ref{boundary}, we additionally discuss the cases where $\alpha_{\rm acc}$ is not uniform.

\subsubsection{Evolution of dust particles} \label{evolution}
Second, we calculate the growth and drift of dust particles. We calculate the peak mass of the particles at each distance from the central planet, $m_{\rm d}$, which is a valid way to investigate the evolution of particles in CPDs \citep{shi17}. We assume that all the dust particles flow into the calculation region at $r=r_{\rm out}$ with $R_{\rm d}=1~{\rm mm}$, where $R_{\rm d}$ is the radius of the particles. The dust mass accretion rate inside the CPDs is
\begin{equation}
\dot{M_{\rm d}}=-2\pi rv_{r}\Sigma_{\rm d},
\label{continuous}
\end{equation}
where $v_{\rm r}$ and $\Sigma_{\rm d}$ are the dust drift speed and the dust surface density. We assume that the dust mass accretion rates inside the CPDs are equal to that flowing onto the CPDs from their parental PPDs. Inside the snowline, $\dot{M_{\rm d}}$ becomes half of the outside, assuming the evaporation of the ice which constitutes the half mass of each particle outside the snowline.

The growth of the drifting particles is calculated by a function of $r$ \citep{sat16},
\begin{equation}
v_{\rm r}\dfrac{d m_{\rm d}}{d r}=\epsilon_{\rm grow}\dfrac{2\sqrt{\pi}R_{\rm d}^{2}\Delta v_{\rm dd}}{H_{\rm d}}\Sigma_{\rm d},
\label{growth}
\end{equation}
where $\epsilon_{\rm grow}$, $\Delta v_{\rm dd}$ and $H_{\rm d}$ are the sticking efficiency for a single collision, collision velocity and dust scale height, respectively. The mass of a single dust particle is $m_{\rm d}=(4\pi/3) R_{\rm d}^{3}\rho_{\rm int}$, where $\rho_{\rm int}=1.4$ and $3.0~{\rm g~cm^{-3}}$ are the internal density of the icy and rocky particles, respectively. In this work, we assume that the particles are compact (see Section~\ref{internaldensity}). We calculate the radial distribution of the peak mass and surface density of the dust particles by equations~(\ref{continuous}) and (\ref{growth}). We substitute equation~(\ref{continuous}) into equation~(\ref{growth}) and integrate it with $r$ from $r=r_{\rm out}$ toward smaller $r$.

The motion of dust particles in gas discs is determined by Stokes number (stopping time normalized by Kepler time) ${\rm St}=t_{\rm stop}\Omega_{\rm K}$. The Stokes number can be written by the following two equations,
\begin{equation}
{\rm St}=
\begin{cases}
\dfrac{\rho_{\rm int}R_{\rm d}}{\rho_{\rm g}v_{\rm th}}\Omega_{\rm K}, & R_{\rm d}\leq\dfrac{9}{4}\lambda_{\rm mfp}, \\
\dfrac{8}{3C_{\rm D}}\dfrac{\rho_{\rm int}R_{\rm d}}{\rho_{\rm g}\Delta v_{\rm dg}}\Omega_{\rm K}, & R_{\rm d}>\dfrac{9}{4}\lambda_{\rm mfp},
\end{cases}
\label{vt}
\end{equation}
where $v_{\rm th}=\sqrt{8/\pi}c_{\rm s}$ is the thermal gas velocity, $\Delta v_{\rm dg}$ is the relative velocity between the dust particles and gas, and $C_{\rm D}$ is a dimensionless coefficient that depends on the particle Reynolds number, ${\rm Re_{p}}$. The upper equation represents the Epstein regime, and the lower equation represents the Stokes and Newton regimes. The particle Reynolds number is
\begin{equation}
{\rm Re_{p}}=\dfrac{4R_{\rm d}\Delta v_{\rm dg}}{v_{\rm th}\lambda_{\rm mfp}},
\label{Rep}
\end{equation}
where $\lambda_{\rm mfp}=m_{\rm g}/(\sigma_{\rm mol}\rho_{\rm g})$ is the mean free path of the gas molecules with their collisional cross section being $\sigma_{\rm mol}=2\times10^{-15}{\rm cm}^{2}$ and with the gas density at the mid-plane being $\rho_{\rm g}=\Sigma_{\rm g}/(\sqrt{2\pi}H_{\rm g})$. We calculate $C_{\rm D}$ as \citep{per11},
\begin{equation}
C_{\rm D}=\dfrac{24}{\rm Re_{p}}\left( 1+0.27{\rm Re_{p}}\right)^{0.43}+0.47\left[1-\exp\left(-0.04{\rm Re_{p}}^{0.38}\right)\right].
\label{CD}
\end{equation}

The dust scale height induced by the vertical diffusion is \citep{you07},
\begin{equation}
H_{\rm d,diff}=H_{\rm g}\left(1+\dfrac{\rm St}{\alpha_{\rm diff}}\dfrac{1+2{\rm St}}{1+{\rm St}}\right)^{-1/2}.
\label{Hddiff}
\end{equation}
The dust scale height is $H_{\rm d}=H_{\rm d,diff}$ unless other effects induce it stronger than the vertical diffusion. We consider the case that the Kelvin-Helmholtz (KH) instability plays a role in Section~\ref{KHI}.

The radial drift velocity of the particles is \citep{whi72,ada76,wei77}
\begin{equation}
v_{\rm r}=-2\dfrac{\rm St}{\rm St^{2}+1}\eta v_{\rm k},
\label{vr}
\end{equation}
where $v_{\rm k}=r\Omega_{\rm k}$ is the Kepler velocity, and
\begin{equation}
\eta=-\dfrac{1}{2}\left(\dfrac{H_{\rm g}}{r}\right)^{2}\dfrac{\partial \ln{\rho_{\rm g}c_{\rm s}^{2}}}{\partial \ln{r}}
\label{eta}
\end{equation}
is the ratio of the pressure gradient force to the gravity of the central planet.

The relative velocity between the dust particles (i.e., collision velocity) is,
\begin{equation}
\Delta v_{\rm dd}=\sqrt{\Delta v_{\rm B}^{2}+\Delta v_{\rm r}^{2}+\Delta v_{\rm \phi}^{2}+\Delta v_{\rm z}^{2}+\Delta v_{\rm t}^{2}},
\label{vdd}
\end{equation}
where $\Delta v_{\rm B}$, $\Delta v_{\rm r}$, $\Delta v_{\rm \phi}$, $\Delta v_{\rm z}$, and $\Delta v_{\rm t}$ are the relative velocities induced by their Brownian motion, radial drift, azimuthal drift, vertical sedimentation, and turbulence, respectively \citep{oku12}. The relative velocity induced by Brownian-motion between the particles with the same mass, $m_{\rm d}$, can be written as $\Delta v_{\rm B}=\sqrt{16k_{\rm B}T/(\pi m_{\rm d})}$. The relative velocity induced by the radial drift is $\Delta v_{\rm r}=|v_{\rm r}({\rm St}_{1})-v_{\rm r}({\rm St}_{2})|$, where ${\rm St}_{1}$ and ${\rm St}_{2}$ are the Stokes numbers of the two colliding particles. The relative velocity induced by the azimuthal drift is $\Delta v_{\rm \phi}=|v_{\rm \phi}({\rm St}_{1})-v_{\rm \phi}({\rm St}_{2})|$, where $v_{\rm \phi}=-\eta v_{\rm K}/(1+{\rm St}^{2})$, and that by the vertical motion is $\Delta v_{\rm z}=|v_{\rm z}({\rm St}_{1})-v_{\rm z}({\rm St}_{2})|$, where $v_{\rm z}=-\Omega_{\rm K}{\rm St}H_{\rm d,diff}/(1+{\rm St})$. We assume ${\rm St}_{2}=0.5{\rm St}_{1}$, because the results of the single size (mass) simulations match those of the full size ones well \citep{sat16}. The relative velocity induced by turbulence (diffusion) is \citep{orm07}
\begin{equation}
\Delta \varv_{\rm t}=
\begin{cases}
\sqrt{\alpha_{\rm diff}}c_{\rm s}{\rm Re}_{\rm t}^{1/4}\left|{\rm St_{1}}-{\rm St_{2}}\right|, & {\rm St_{1}}\ll {\rm Re}_{\rm t}^{-1/2}, \\
\sqrt{3\alpha_{\rm diff}}c_{\rm s}{\rm St}_{1}^{1/2}, & {\rm Re}_{\rm t}^{-1/2} \ll {\rm St_{1}}\ll 1, \\
\sqrt{\alpha_{\rm diff}}c_{\rm s}\left(\dfrac{1}{1+{\rm St_{1}}}+\dfrac{1}{1+{\rm St_{2}}}\right)^{1/2}, & 1\ll {\rm St_{1}}.
\end{cases}
\label{St}
\end{equation}
The turbulence Reynolds number is ${\rm Re_{t}}=\nu/\nu_{\rm mol}$, where $\nu_{\rm mol}=v_{\rm th}\lambda_{\rm mfp}/2$ is the molecular viscosity. We calculate the relative velocity between the particles and gas, $\Delta v_{\rm dg}$, by setting ${\rm St_{1}}={\rm St}$ and ${\rm St_{2}}\rightarrow0$ in the above expressions.

When the collision speed is too high, the particles break up rather than merge. Previous numerical simulations show that the critical velocity for the collision of the rocky and icy particles are about $v_{\rm cr}=5$ and $50~{\rm m~s^{-1}}$, respectively \citep{wad09,wad13}. The sticking efficiency for a single collision is written as,
\begin{equation}
\epsilon_{\rm grow}=\min\left\{1, -\dfrac{\ln{(\Delta v_{\rm dd}/v_{\rm cr})}}{\ln{5}}\right\},
\label{stfrag}
\end{equation}
from the fitting of the simulations \citep{oku16}. When the collision velocity, $\Delta v_{\rm dd}$, is faster than the critical velocity, $v_{\rm cr}$, the size of the particles is determined by the fragmentation (see also Section~\ref{feasibility}).

\subsection{Results} \label{results}
\subsubsection{Fiducial case} \label{fiducial}
\begin{figure*}
\centering
\includegraphics[width=\linewidth]{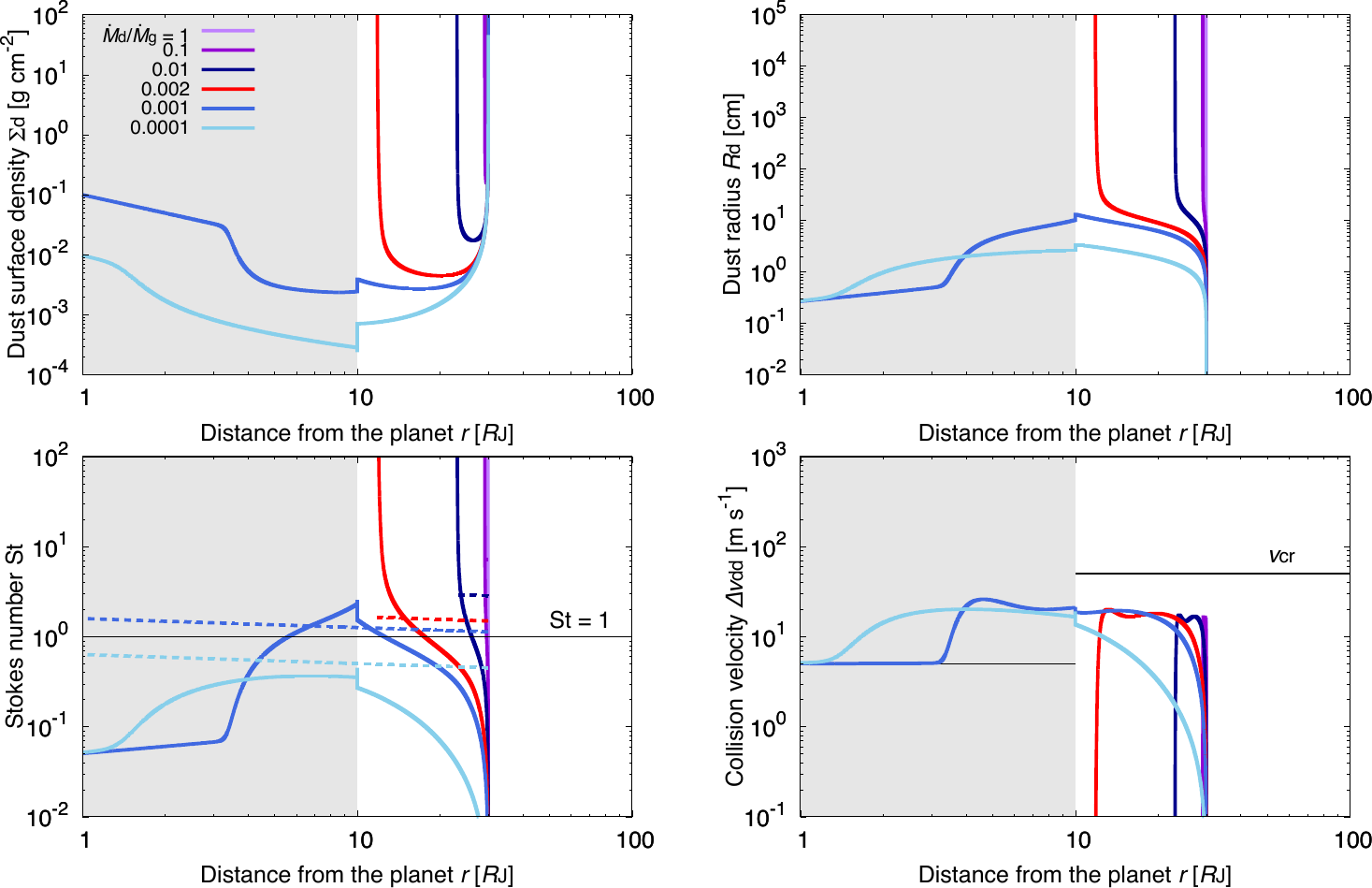}
\caption{Evolution of dust particles in a CPD with $\alpha_{\rm acc}=3.1\times10^{-3}$, $\alpha_{\rm diff}=7.4\times10^{-8}$ and $\dot{M}_{\rm g}=0.15~M_{\rm J}~{\rm Myr}^{-1}$. The upper left, upper right, lower left and lower right panels represent the radial distribution of the surface density, radius, Stokes number and collision velocity of the dust particles, respectively. The colour variation of the curves represents the difference of $\dot{M}_{\rm d}/\dot{M}_{\rm g}$. Especially, the red curve represents the evolution with $\dot{M}_{\rm d}/\dot{M}_{\rm g}=0.002$, the critical value of the condition for the satellitesimal formation. In the lower left panel, the dashed lines are the analytical estimates given by equation~(\ref{Stalpha}) and the black horizontal line represents ${\rm St}=1$. The black horizontal lines in the lower right panel are the critical velocity of fragmentation. The shaded grey is the region inside the snowline, where $T\geq160~{\rm K}$.}
\label{fig:evolution}
\end{figure*}

We calculate the evolution of dust particles in the CPD with the various $\dot{M}_{\rm d}/\dot{M}_{\rm g}$ taking the value from $0.0001$ to $1$. Figure~\ref{fig:evolution} shows that the particles grow at their inflowing point and, if $\dot{M}_{\rm d}/\dot{M}_{\rm g}\leq0.01$, they start to drift when their sizes become $\sim1-10~{\rm cm}$ (i.e., pebble-size). The figure also shows that the particles grow rapidly if the Stokes number reaches unity. The drift velocity $v_{\rm r}$ is fastest when ${\rm St}=1$ (see equation~(\ref{vr})). Therefore, if the dust particles grow and the Stokes number becomes larger than unity, the particles start to pile up and increase the dust density, resulting in their rapid growth to satellitesimals.

This local condition for the rapid growth in the CPD, ${\rm St} \gtrsim 1$, is the same with that for planetesimal formation in PPDs. For example, \citet{tak21} calculated the growth of dust by time-evolving gas and dust distributions in PPDs and found that the condition for the dust accumulation is $\Sigma_{\rm d}/\Sigma_{\rm g}\gtrsim\eta$ (Eq.~(28) in the paper). However, this condition cannot be applied to our work directly. This is because the equation is for the dust particles in the Epstein regime, but they are mainly in the Stokes or Newton regimes in CPDs \citep{shi17}, Plus, in the case of the dust evolution in CPDs, $\dot{M}_{\rm d}/\dot{M}_{\rm g}$ is a more essential parameter than $\Sigma_{\rm d}/\Sigma_{\rm g}$, because the dust and gas are continuously supplied to the CPDs, and quasi-steady states are realized inside the discs\footnote{\citet{tak21} shows another local condition for the dust accumulation, ${\partial\dot{M}_{\rm d}}/{\partial r}<0$ (Eq.~(30) in the paper). However, we assume that the radial dust flux is uniform, so it is not possible to apply the condition to our work directly either.}.

Figure~\ref{fig:evolution} also shows that satellitesimals can form when $\dot{M}_{\rm d}/\dot{M}_{\rm g}\geq0.002$. This critical value is $500$ times smaller than that in the previous work, $\dot{M}_{\rm d}/\dot{M}_{\rm g}\geq1$ \citep{shi17}. In the case of $\dot{M}_{\rm d}/\dot{M}_{\rm g}=0.001$, the Stokes number becomes larger than unity just outside the snowline, but the particles flow into the inside of the snowline before they grow to satellitesimals. Inside the snowline, the fragmentation of the particles occurs, and thus the dust size keeps small ($\lesssim10~{\rm cm}$). This is because the particles are dry inside the snowline, and the critical fragmentation velocity is slow, so that the collision velocity $\Delta v_{\rm dd}$ is faster than the critical velocity (see the lower right panel of Fig.~\ref{fig:evolution}). The small increase of the Stokes number at the snow line is because the internal density is larger inside the snowline, whose impact is larger than the decrease of the pebble mass flux. With $\dot{M}_{\rm d}/\dot{M}_{\rm g}=0.0001$, the particles grow slower than those with $\dot{M}_{\rm d}/\dot{M}_{\rm g}=0.001$. In the snowline, the size is determined by the fragmentation similar to that in the case with $\dot{M}_{\rm d}/\dot{M}_{\rm g}=0.001$.

These results can be understood by the following analytical approximation of the Stokes number when the dust size is determined by its drift, in other words, when it is outside the snowline,
\begin{equation}
\begin{split}
{\rm St}\approx&1.7\left(\dfrac{\dot{M}_{\rm d}/\dot{M}_{\rm g}}{0.002}\right)^{2/5}\left(\dfrac{\alpha_{\rm diff}}{7.4\times10^{-8}}\right)^{-1/5}\left(\dfrac{\alpha_{\rm acc}}{3.1\times10^{-3}}\right)^{2/5} \\
&\times\left(\dfrac{T}{160~{\rm K}}\right)^{-2/5}\left(\dfrac{M_{\rm p }}{1~M_{\rm J}}\right)^{2/5}\left(\dfrac{r}{10~R_{\rm J}}\right)^{-2/5},
\label{Stalpha}
\end{split}
\end{equation}
which is derived by integrating equation~(\ref{growth}) from an outer radius to $r$ and taking $r$ to be much smaller than the outer radius. When the power-law indices of the $r$ dependency of $\Sigma_{\rm g}$ and $T$ ($\Sigma_{\rm g}\propto r^{-p}$ and $T\propto r^{-q}$) are parameters, the numerical factor $1.7$ depends on the indices, $p$ and $q$, the factor is $3.4\times(2/(3+2p+q))^{4/5}\times(10/(18-13q))^{2/5}$, which can be used for more general cases. In Eq.(\ref{Stalpha}), the Stokes number depends on the dust-to-gas mass ratio of the inflow, $\dot{M}_{\rm d}/\dot{M}_{\rm g}$, and is larger than unity when $\dot{M}_{\rm d}/\dot{M}_{\rm g}\geq0.002$ at the snowline ($r=10~R_{\rm J}$ and $T=160~{\rm K}$). Therefore, when $\dot{M}_{\rm d}/\dot{M}_{\rm g}\geq0.002$, the Stokes number of the dust can reach unity with an enough margin before the particles cross the snowline, and satellitesimals form. The dashed lines in the left bottom panel of Fig.~\ref{fig:evolution} represent the approximations.

\subsubsection{Effects of the Kelvin-Helmholtz instability} \label{KHI}
In reality, the KH instability induces turbulence keeping the dust scale height at some height \citep[e.g.][]{chi10}. The dust scale height induced by the KH instability can be written as,
\begin{equation}
\begin{split}
H_{\rm d,KH}&={\rm Ri}^{1/2}\dfrac{Z^{1/2}}{(1+Z)^{3/2}}\eta r \\
&=\{{\rm Ri}Z_{\rm \Sigma}H_{\rm g}(\eta r)^{2}\}^{1/3} - Z_{\rm \Sigma}H_{\rm g}
\label{HdKH}
\end{split}
\end{equation}
where ${\rm Ri}$ is the Richardson number for the particles, which is about  $0.5$ when the KH instability supports the dust layer \citep{hyo21a}. The dust-to-gas mid-plane density ratio and the dust-to-gas surface density ratio are $Z=\rho_{\rm d}/\rho_{\rm g}$ and $Z_{\rm \Sigma}=\Sigma_{\rm d}/\Sigma_{\rm g}$, respectively. The KH instability gives the minimum dust scale height when the particles are small. On the other hand, the KH instability does not grow when they are large \citep{mic06}. Thus, we simply assume that the instability only occurs when ${\rm St}<1$. We calculate the scale height as,
\begin{equation}
H_{\rm d}=
\begin{cases}
\max\{H_{\rm d,diff}, H_{\rm d,KH}\}, & {\rm St}<1, \\
H_{\rm d,diff}, & 1\leq{\rm St}.
\end{cases}
\label{Hd}
\end{equation}
where $H_{\rm d,diff}$ is the dust scale height induced by the vertical diffusion (equation~(\ref{Hddiff})). The KH instability may also change the collision velocity $\Delta v_{\rm t}$ of equation~(\ref{vdd}), but it can be ignored because $\Delta v_{\rm r}$ is much larger. We check this by calculating the collision velocity $\Delta v_{\rm t}$ induced by the KH instability, by solving equation~(\ref{Hddiff}) for $\alpha_{\rm diff}$ and using equation~(\ref{vt})\footnote{Also as an order-of-magnitude estimate, when ${\rm St}=1$, the ratio of two velocity is $\Delta v_{\rm t}/\Delta v_{\rm r}\sim\sqrt{\alpha_{\rm diff}}/h_{\rm g}\sim H_{\rm d,KH}/H_{\rm g}\times h_{\rm g}^{-1}\sim0.1$, where $H_{\rm d,KH}/H_{\rm g}\approx0.005$ and the aspect ratio $h_{\rm g}=H_{\rm g}/r\approx0.05$ (see Fig.~\ref{fig:Hp}).}.

\begin{figure*}
\centering
\includegraphics[width=\linewidth]{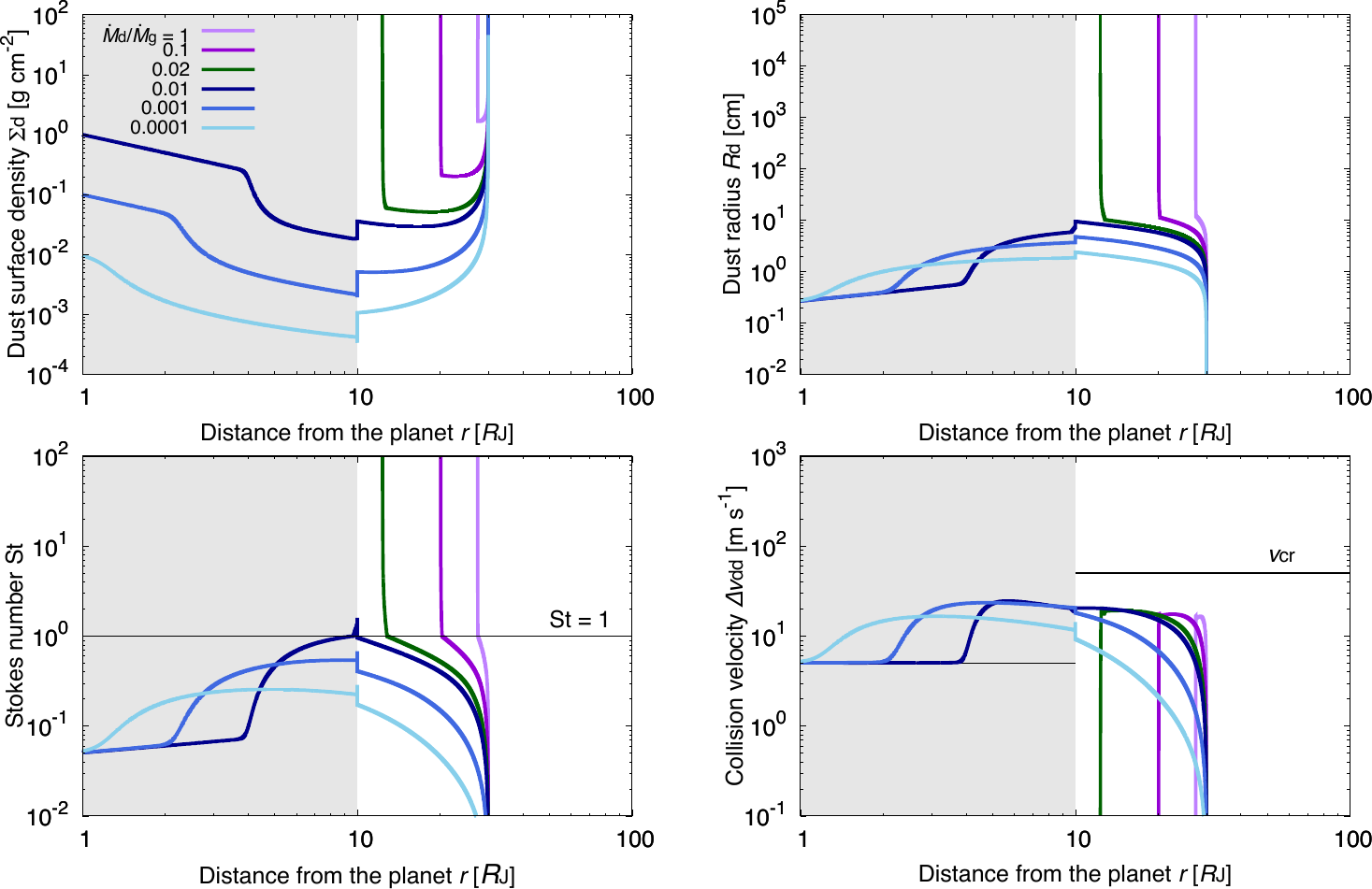}
\caption{Same as Fig.~\ref{fig:evolution} but with the KH instability. The dark green curve represents the evolution with $\dot{M}_{\rm d}/\dot{M}_{\rm g}=0.02$, the critical value of the condition for the satellitesimal formation.}
\label{fig:KHI}
\end{figure*}

\begin{figure}
\centering
\includegraphics[width=\columnwidth]{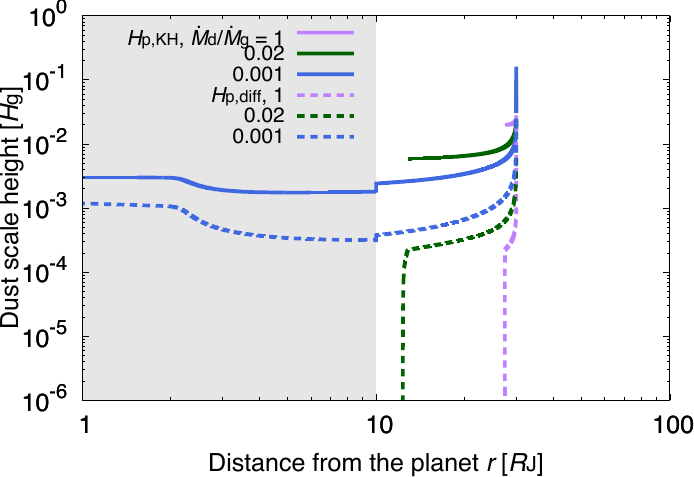}
\caption{Dust scale height induced by the KH instability (solid) and the vertical diffusion (dotted). The profiles are normalized by the gas scale height. The purple, dark green and blue curves represent the profiles with $\dot{M}_{\rm d}/\dot{M}_{\rm g}=1$, $0.02$ and $0.001$, respectively.}
\label{fig:Hp}
\end{figure}

Figure~\ref{fig:KHI} shows that the condition for the satellitesimal formation with the effect of the KH instability is $\dot{M}_{\rm d}/\dot{M}_{\rm g}=0.02$, which is ten times larger than the critical value without the effect. This is because the dust scale height becomes higher by the KH instability, and the collision rate is proportional to its inverse (equation~(\ref{growth})). Fig.~\ref{fig:Hp} shows that $H_{\rm d,diff}$ is about more than ten times larger than $H_{\rm d,diff}$ outside the snowline in any case.

\section{Discussion}\label{discussion}
\subsection{Estimate of the wind-launching region} \label{region}
In Section~\ref{MHD}, we do not set the vertical gas flowing into the simulation boxes in our MHD simulations. The vertical inflow should continuously accrete from the high altitude to the surfaces of CPDs, which may suppress the launch of the disc wind. In this section, we estimate the region where the disc wind can be launched against the gas inflow by comparing the magnetic pressure generated in the disc with the Ram pressure. The disc wind is driven by the magnetic pressure of the toroidal field, while the Ram pressure of the inflow should be the pressure suppressing the wind. We note that this comparison can only provide an evaluation of the situation after the disc wind launches once. The gas inflow has to be calculated directly by MHD simulations to investigate whether the disc wind can launch at its first time (see also Section~\ref{comparison}).

Table~\ref{tab:MHD} represents the toroidal magnetic pressure $ P_{\rm mag}=B_{y}^2/(8 \pi)$ measured at $z=4H_{\rm g}$, where the disc wind is launched, in each simulation. The values are normalized by the gas pressure at the mid-plane, $P_{\rm mid}=\Sigma_{\rm g} c_{\rm s} \Omega_{\rm K}/\sqrt{2 \pi}$, assuming that the vertical gas density profile is in the hydrostatic equilibrium.

We evaluate the Ram pressure by assuming that the kinetic energy of the inflow is conserved over the path from the PPD to the CPD. The inflow Ram pressure is $P_{\rm Ram}=\rho_{\rm inf}v_{\rm inf}^2 $, where $\rho_{\rm inf}$ and $v_{\rm inf}$ are the gas density and velocity of the inflow, respectively. The inflow velocity can be approximated as the free fall velocity \citep{tan12}, $v_{\rm inf}\approx\sqrt{2GM_{\rm p}/r}$. Here, we assume that the mass accretion rate of the CPD is equal to that of gas inflow onto the CPD from the PPD, $\dot{M}_{\rm g} = 2\pi R_{\rm inf}^{2} \rho_{\rm inf}v_{\rm inf}$, where $R_{\rm inf}$ is the outer edge of the gas inflow region. We obtain
\begin{equation}
\frac{P_{\rm Ram}}{P_{\rm mid}}=\frac{3\sqrt{\pi}\alpha_{\rm acc}H_{\rm g}r}{R_{\rm inf}^{2}},
\label{Pram}
\end{equation}
using the above equations and equation~(\ref{alpha-acc}). We also assume that the gas inflow flux is uniform in the region and $R_{\rm inf}=0.1R_{\rm H}$, where $R_{\rm H}\equiv(M_{\rm p}/(3M_{*}))^{1/3}a_{\rm p}\approx1000~R_{\rm J}$ is the Hill radius of the planet (see Fig.~13 of \citet{tan12}). The mass of the central star is $M_{*}=1M_{\odot}$ (one stellar mass) and the orbital distance of the planet is $a_{\rm p}=5.2~{\rm au}$.

\begin{figure}
\centering
\includegraphics[width=\columnwidth]{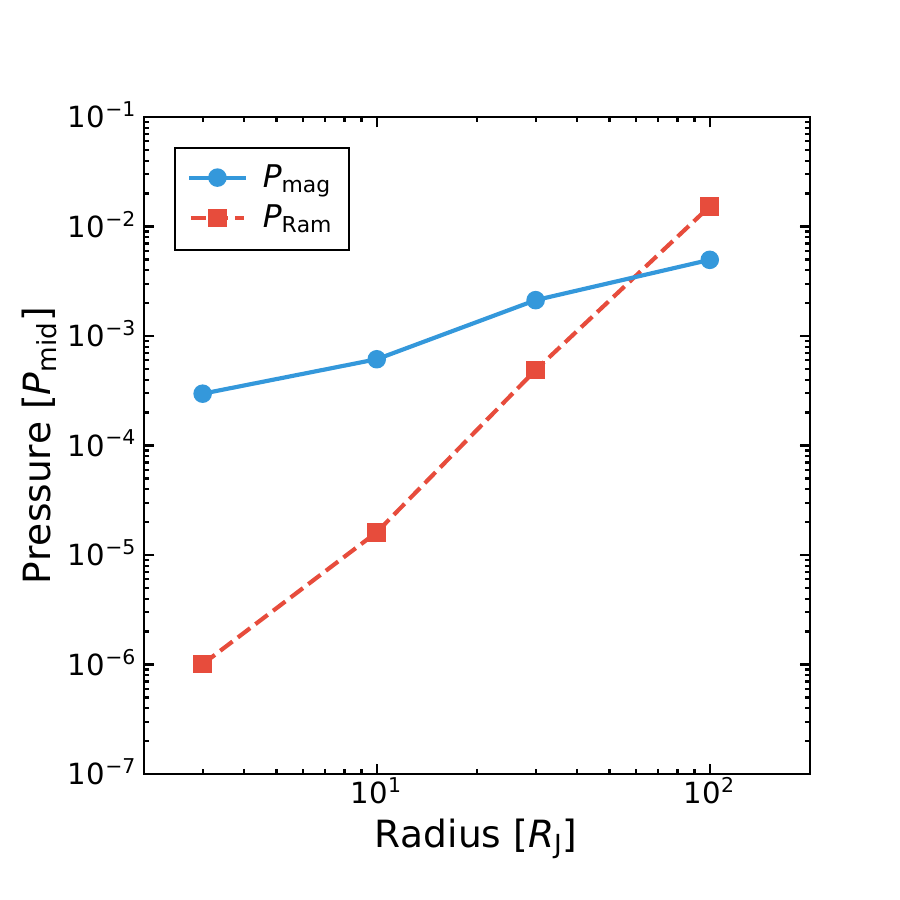}
\caption{Magnetic pressure $P_{\rm mag}$ (blue solid) and Ram pressure $P_{\rm Ram}$ (red dashed) as a function of distance from the planet. The magnetic pressure is measured at $z=4H_{\rm g}$ where the disc wind is launched. The Ram pressure is assumed to be equal to the gravitational potential at the radius.}
\label{fig:pmag-prof}
\end{figure}

Figure~\ref{fig:pmag-prof} represents the magnetic pressure and the Ram pressure at each radius. We find that the magnetic pressure is larger than the Ram pressure within $r\approx30~R_{\rm J}$. This suggests that the magnetic disc wind is expected to be launched inside 30 $R_{\rm J}$, whereas the gas inflow from PPDs accretes onto the CPD and the wind is suppressed beyond the radius.
This suggests that the wind-accretion CPDs shown in this work can be stable with preventing the gas inflow from the PPD by the wind pressure. 
Direct verification of this expectation by numerical simulations would be a future work.

\subsection{Comparison of the MHD simulations to previous works} \label{comparison}
The obtained accreting rate, $\dot{M}_{\rm g}\sim0.2~M_{\rm J}~{\rm Myr}^{-1}$, is $\sim100$ times smaller than that shown in most of the previous (M)HD simulations considering the full gap opening \citep[e.g.,][]{gre13,lam19,sch19b,sch20}. What does this mean?

First, the value of the gas accretion rate obtained by our MHD simulation can be changed when we give another gas surface density of the CPD as the initial setting. We could reproduce much higher gas accretion rate by giving a higher gas surface density, but it is not the purpose of this paper. Our purpose is to reproduce a suitable condition for the satellitesimal formation. Especially, the temperature must be lower than the sublimation-point of water to form icy satellites(imals). Such low temperature can be achieved only when the gas accretion rate is smaller than $\sim0.2~M_{\rm J}~{\rm Myr}^{-1}$ (see Figure 1 of \citet{shi17}). Actually, most of the previous works of the satellite formation also assumed the same or smaller gas accretion rate \citep[e.g.,][]{can02}.

The obtained smaller gas accretion rate in the CPD compared to the gas inflow rate in the other previous (M)HD simulations can be interpreted that our simulations reproduce a different phase of the gas accretion of the planet from those of the previous works. The previous simulations consider the main phase of the PPD evolution. On the other hand, our simulations reproduce a later phase where the gas inflow rate has decreased by the dissipation of the PPD gas.

\citet{gre13} calculated global MHD simulations of a CPD including the gas inflow from the PPD, which should be compared with our work as important MHD simulation research of CPDs. Although MRI is active at the surfaces of the disc in their simulation, it is dead in the whole height of the disc in our simulations. This is because ambipolar diffusion, a nonideal MHD effect not included in the previous simulation, suppresses MRI in the higher regions than the mid-plane. \citet{gre13} also found that the gas accretion inflows onto CPDs have temporal variation, especially when the magnetic field plays an important role. This suggests the estimate in Section~\ref{region} could be inadequate because we implicitly assumed the gas accretion rate is constant. Additionally, \citet{gre13} found that jets can be launched episodically, which may change the structures of the gas and magnetic field especially of the inner parts of the CPDs. However, the phase considered in the work should be different from that of this work, as we discussed above. It is not clear if the jet can also be launched in the satellite formation phase, where the gas accretion rate is much lower than that in \citet{gre13}. It is difficult to discuss the jet by local shearing box simulations. In any case, global MHD simulations including all the three nonideal effects with much longer timescale should be carried out, but it is beyond the scope of this paper.

Two-dimensional high resolution HD simulations by \citet{zhu16} showed that spiral shocks (arms) can also contribute to the angular momentum transport of CPDs and could be the dominant accretion mechanism. The accretion efficiency by this mechanism with one of the simulation settings is $\alpha_{\rm acc}\sim10^{-3}$--$10^{-2}$ with $\dot{M}_{\rm g}\sim0.1$--$1~M_{\rm J}~{\rm Myr}^{-1}$, which is the same order-of-magnitude with that of the magnetic wind-driven accretion predicted in this work. Again, 3D global nonideal MHD simulations are needed to investigate which accretion mechanism is dominant.

\subsection{Feasibility of the satellitesimal formation} \label{feasibility}
In Section~\ref{results}, we obtained the condition for the satellitesimal formation with the effect of the KH instability, $\dot{M}_{\rm d}/\dot{M}_{\rm g}\geq0.02$. This required value is much smaller than that of the previous work. The next question is whether it is possible to fulfill the condition.

\citet{hom20} found that if there are no gas gaps around the CPDs, almost all dust particles around there can flow into the CPDs. In the case that the gas around the CPDs has solar composition, i.e, the dust-to-gas surface density ratio is $0.01$, the dust-to-gas mass ratio of the inflow to the CPDs also has almost the same value, $\dot{M}_{\rm d}/\dot{M}_{\rm g}\approx0.01$ which is comparable to the critical value. Therefore, satellitesimals may be able to form in this case.

On the other hand, the gap has opened in the situation considered in this work (see Section~\ref{comparison}), and so the filtering effect by the gap is a problem. Relatively large dust particles (i.e., pebbles drifting from the outer region of PPDs) cannot reach the inside of the gap due to the gas pressure at the edge of the gap \citep[e.g.][]{paa04}. The dust-to-gas mass ratio of the inflow gas is significantly reduced by this filtering effect \citep{hom20}.

However, dust particles are piled up at the gas pressure bump, and so the dust-to-gas density ratio dramatically increases at the bump \citep[e.g.][]{kan18}, resulting in formation of a lot of small fragments by the mutual collision of the dust particles \citep{dra19}. Moreover, the filtering effect strongly depends on the strength of diffusion at the outer edge of the gap; strong turbulent diffusion strews the particles over the inside of the gap against the gas pressure gradient \citep[e.g.][]{zhu12,shi20}. Once the particles reach the inside of the gap, most of them are blown up by turbulence (if it is not so week inside the gap) and are supplied to the CPDs with the vertically inflowing gas \citep{hom20}, which may fulfill the condition for the satellitesimal formation. Furthermore, recent 3D global dust+gas radiative hydrodynamic simulations by \citet{szu21} show that the ``meridional circulation'', a larger scale three-dimensional gas flow bridging the gap, can directly deliver mm-sized particles onto the CPD even from the outside of the gap. In the case of a Jupiter mass planet, $\sim0.01~M_{\rm J}~{\rm Myr}^{-1}$ of dust can be supplied to the vicinity ($0.5~R_{\rm H}$) of the planet with the gas flow of $\sim0.1~M_{\rm J}~{\rm Myr}$, in other words, $\dot{M}_{\rm d}/\dot{M}_{\rm g}\sim0.1$. Therefore, satellitesimals can form by our scenario even if the effects of the KH instability is included.

Dust fragmentation in CPDs could be a barrier for the satellitesimal formation. We set the critical velocity of rocky and icy particles as $v_{\rm cr}=5$ and $50~{\rm m~s^{-1}}$, respectively, according to the results of numerical simulations assuming the monomer size as $0.1\mu {\rm m}$ \citep{wad09,wad13}. However, the theory of dust sticking predicts that the critical velocity scales with the $-5/6$ power of the monomer size of the particles \citep[e.g.][]{cho93}, which is consistent with the results of the experiments of collision using $\sim\mu {\rm m}$-sized particles, $v_{\rm cr}=1$ and $10~{\rm m~s^{-1}}$ \citep{blu00,pop00,gun15}. Thus, if the monomer size of dust supplied to CPDs, which must be the same with that in PPDs, is $\sim\mu {\rm m}$, satellitesimals cannot form due to the fragmentation (see the right bottom panels of Figs~\ref{fig:evolution} and \ref{fig:KHI}). We note that, however, the typical size of the interplanetary dust particles of presumably cometary origin is $\sim0.1\mu {\rm m}$ \citep[e.g.][]{rie93}, which may be the monomer size of the dust in PPDs.

\subsection{Dependence of turbulent strength and accretion efficiency} \label{alphaspace}
Although the results of our nonideal MHD simulations show that $\alpha_{\rm acc}=3.1\times10^{-3}$ and $\alpha_{\rm diff}=7.4\times10^{-8}$ (Section~\ref{wind-driven}), these results should depend on the assumptions. Thus, we calculate the evolution of the dust particles with various sets of $\alpha_{\rm acc}$ and $\alpha_{\rm diff}$, by varying the two parameters from $10^{-8}$ to $10^{-2}$.

\begin{figure}
\centering
\includegraphics[width=\columnwidth]{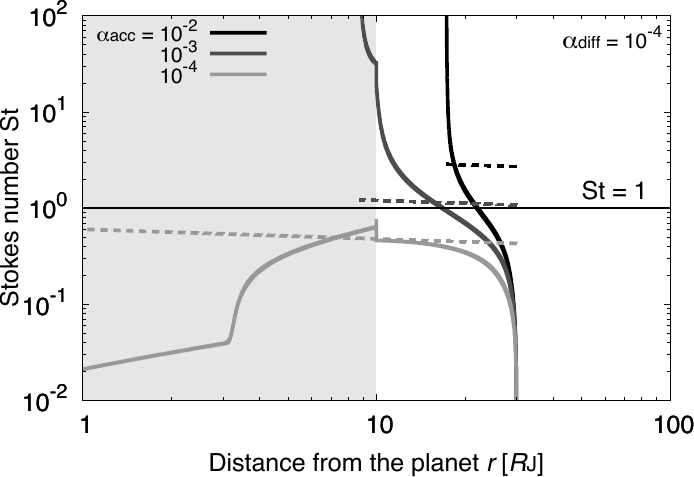}
\caption{Same as the lower left panel of Fig.~\ref{fig:evolution} but for various $\alpha_{\rm acc}$ and $\dot{M}_{\rm d}/\dot{M}_{\rm g}=0.1$. The grayscale of the curves represents the difference of $\alpha_{\rm acc}$. The strength of diffusion is fixed as $\alpha_{\rm diff}=10^{-4}$.}
\label{fig:alphaacc}
\end{figure}

Figure~\ref{fig:alphaacc} represents the results of the cases where $\alpha_{\rm diff}$ is fixed to be $10^{-4}$, and $\alpha_{\rm acc}$ is changed. Satellitesimals can form when $\alpha_{\rm acc}=10^{-3}$ and $10^{-2}$, whereas the smaller $\alpha_{\rm acc}$ is, the smaller dust particles drift toward the central planet. The larger $\alpha_{\rm acc}$ results in the lower gas surface density, leading to the slow drift of the particles. This is consistent with the $\alpha_{\rm acc}$ dependence of equation~(\ref{Stalpha}). Moreover, equation~(\ref{Stalpha}) is also obtained by distinguishing the $\alpha$ dependence of equation~(15) of \citet{shi17} into $\alpha_{\rm acc}$ and $\alpha_{\rm diff}$ dependencies,
\begin{equation}
\begin{split}
{\rm St}\approx&1.2\left(\dfrac{\dot{M}_{\rm d}/\dot{M}_{\rm g}}{1}\right)^{2/5}\left(\dfrac{\alpha_{\rm diff}}{10^{-4}}\right)^{1/5}\left(\dfrac{\alpha_{\rm acc}}{\alpha_{\rm diff}}\right)^{2/5} \\
&\times\left(\dfrac{T}{160~{\rm K}}\right)^{-2/5}\left(\dfrac{M_{\rm p }}{1~M_{\rm J}}\right)^{2/5}\left(\dfrac{r}{10~R_{\rm J}}\right)^{-2/5}.
\label{Stalphab}
\end{split}
\end{equation}
The increase in $\alpha_{\rm acc}/\alpha_{\rm diff}$ is equivalent to that in $\dot{M}_{\rm d}/\dot{M}_{\rm g}$ in this estimate. Therefore, even when $\dot{M}_{\rm d}/\dot{M}_{\rm g}=0.1$, which is ten times smaller than the critical value of satellitesimal formation in the previous work ($\dot{M}_{\rm d}/\dot{M}_{\rm g}\geq1$ when $\alpha_{\rm acc}/\alpha_{\rm diff}=1$), the estimated Stokes number can be roughly the same with the previous one and reach unity, resulting in the satellitesimal formation.

\begin{figure}
\centering
\includegraphics[width=\columnwidth]{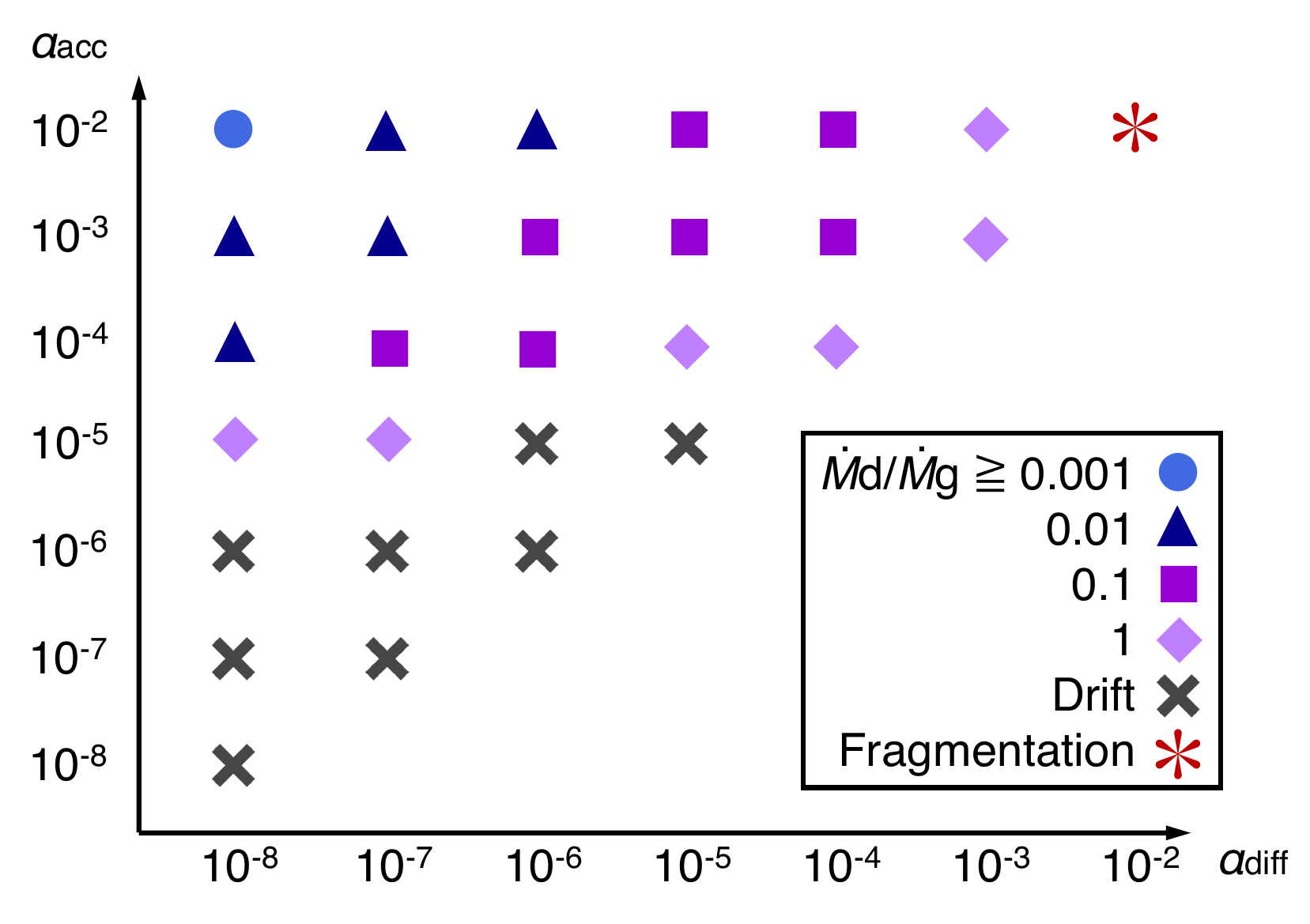}
\caption{Conditions for the satellitesimal formation with different set of $\alpha_{\rm acc}$ and $\alpha_{\rm diff}$. The colour (and symbol) variation indicates the minimum $\dot{M}_{\rm d}/\dot{M}_{\rm g}$ conditions for the satellitesimal formation. The dust particles drift into the planet at the black cross cases. The asterisk indicates the occurrence of the fragmentation of the particles outside the snowline. The gas accretion rate is $\dot{M}_{\rm g}=0.15~M_{\rm J}~{\rm Myr}^{-1}$.}
\label{fig:condition_alpha}
\end{figure}

Figure~\ref{fig:condition_alpha} represents the obtained conditions of $\dot{M}_{\rm d}/\dot{M}_{\rm g}$ that satellitesimals can form in $\alpha_{\rm acc}$--$\alpha_{\rm diff}$ space. In the case of $\alpha_{\rm diff}=10^{-8}$, satellitesimals can form even if the dust-to-gas mass ratio of the inflow onto the CPDs is very small, which is $\dot{M}_{\rm d}/\dot{M}_{\rm g}=0.0001$ when $\alpha_{\rm acc}=10^{-2}$. On the other hand, in the case of strong $\alpha_{\rm diff}$, the condition for the satellitesimal formation is stricter than that with weak $\alpha_{\rm diff}$. The fragmentation of the icy particles can be a problem when $\alpha_{\rm diff}$ is strong, $\alpha_{\rm diff}=10^{-2}$. The correlation between the required values of $\alpha_{\rm acc}$ and $\alpha_{\rm diff}$ for the satellitesimal formation is about $\alpha_{\rm acc}\propto\alpha_{\rm diff}^{1/2}$, which is consistent with equation~(\ref{Stalpha}). We note that the diagonal results show the cases with $\alpha_{\rm acc}$=$\alpha_{\rm diff}$ and are consistent with the results of the previous work by \citet{shi17} (see its Fig.~6).

\subsection{Effects of the internal density of dust particles} \label{internaldensity}
In this work, we assume that the particles are compact and fix the internal density as $\rho_{\rm int}=1.4~{\rm g~cm^{-3}}$ outside the snowline. However, the particles could be porous (low internal density), which may change the growth and drift of them. In PPDs, the internal density of the icy particles can be $\sim10^{-5}$-$10^{-4}~{\rm g~cm^{-3}}$, resulting in the icy planetesimal formation against the drift barrier \citep{oku12}. Here, we investigate the evolution of the particles with $\rho_{\rm int}=1.4\times10^{-1}$, $1.4\times10^{-2}$ and $1.4\times10^{-4}~{\rm g~cm^{-3}}$. The other parameters are the same with those in Section~\ref{KHI}.

\begin{figure}
\centering
\includegraphics[width=\columnwidth]{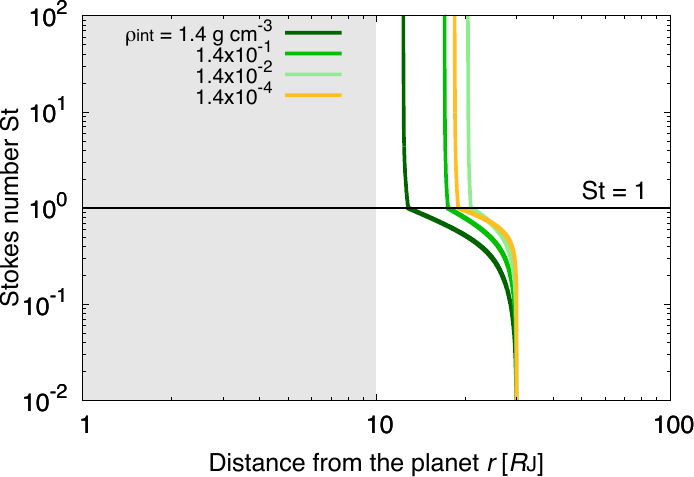}
\caption{Same as the lower left panel of Fig.~\ref{fig:evolution}, but for various value of $\rho_{\rm int}$ and with the KH instability. The colour variation of the curves represent the difference of $\rho_{\rm int}$. The dust-to-gas mass ratio of the inflow is $\dot{M}_{\rm d}/\dot{M}_{\rm g}=0.02$.}
\label{fig:rhoint}
\end{figure}

Figure~\ref{fig:rhoint} shows that the condition for the satellitesimal formation does not change much by the difference of the internal density of dust particles. When $\rho_{\rm int}\ge1.4\times10^{-2}~{\rm g~cm^{-3}}$, the position where satellitesimals form becomes more outside of the CPD as the internal density is smaller. Therefore, the condition for the satellitesimal formation, $\dot{M}_{\rm d}/\dot{M}_{\rm g}\ge0.02$, obtained in Section~\ref{KHI} represents the strictest case. On the other hand, the particles with $\rho_{\rm int}=1.4\times10^{-4}~{\rm g~cm^{-3}}$ do not grow faster than those with $\rho_{\rm int}=1.4\times10^{-2}~{\rm g~cm^{-3}}$. This is because the particles with $\rho_{\rm int}\ge1.4\times10^{-2}~{\rm g~cm^{-3}}$ are in the Stokes regime (${\rm Re_{p}}\lesssim10^{3}$), but with $\rho_{\rm int}=1.4\times10^{-4}~{\rm g~cm^{-3}}$, the particles quickly move into in the Newton regime (${\rm Re_{p}}\gtrsim10^{3}$). The evolution of particles depends on the internal density when they are in the Stokes regime, and this dependence helps the rapid growth of particles, but the dependence disappears in the Newton regime \citep{oku12}.
 
\subsection{Effects of non-uniform efficiency of angular momentum transport of the CPD}
\label{non-uniform}
\begin{figure}
\centering
\includegraphics[width=\columnwidth]{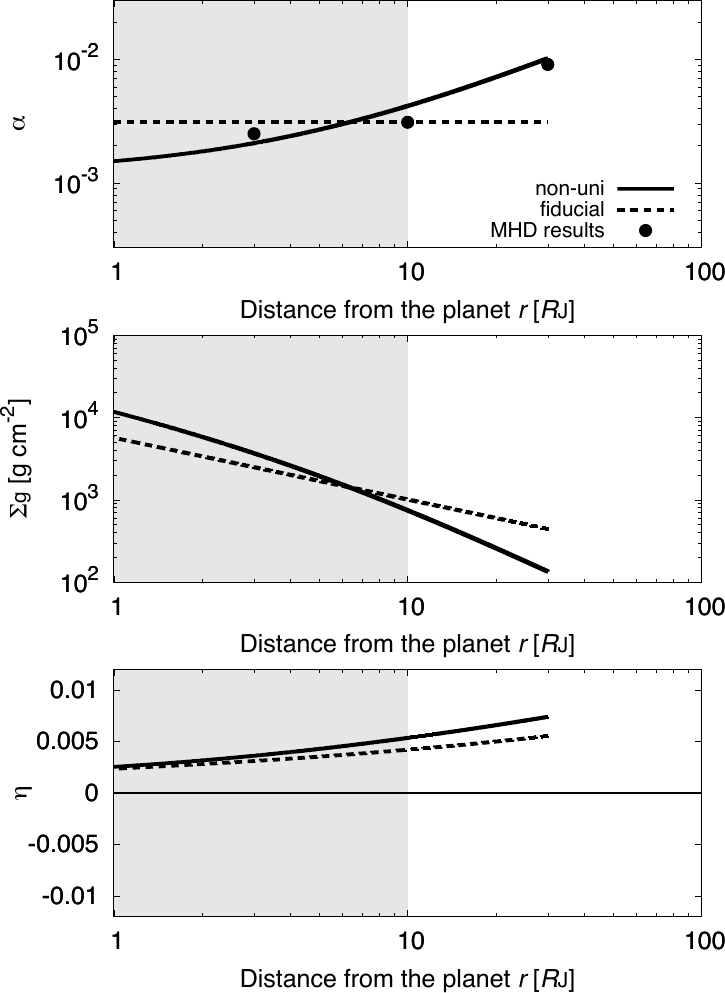}
\caption{Radial distribution of the efficiency of the angular momentum $\alpha_{\rm acc}$, the gas surface density $\Sigma_{\rm g}$, and the ratio of the pressure gradient force to the gravity of the planet $\eta$. The dotted curves represent the profiles of the fiducial uniform $\alpha_{\rm acc}$ case ($\alpha_{\rm acc}=3.1\times10^{-3}$). The black points in the top panel represent the $\alpha_{\rm acc}$ obtained by the MHD simulations.}
\label{fig:alpha-Sigma-eta_non-uni}
\end{figure}

We assume that the efficiency of angular momentum transport of the CPD, $\alpha_{\rm acc}$, is uniform in this work except for the transition region modelled in Section \ref{boundary}. We here consider a case that $\alpha_{\rm acc}$ is not uniform but $\dot{M}_{\rm g}$ is still uniform. We approximate the results of the MHD simulations at $3$, $10$, and $30~R_{\rm J}$\footnote{The disc wind can launch only at $r\leq30~R_{\rm J}$ according to our estimate (Section \ref{region}), so the value at $r=100~R_{\rm J}$ should not be used for the approximation.} by a simple linear function,
\begin{equation}
\alpha_{\rm acc}=3\times10^{-4}\left(\dfrac{r}{R_{\rm J}}\right) + 1.2\times10^{-3}.
\label{alpha_non-uni}
\end{equation}
In this case, the radial profiles of the gas surface density is different from the fiducial model used in  the dust evolution and the assumed setting in the MHD simulations. Figure~\ref{fig:alpha-Sigma-eta_non-uni} shows the difference between the models.

\begin{figure}
\centering
\includegraphics[width=\columnwidth]{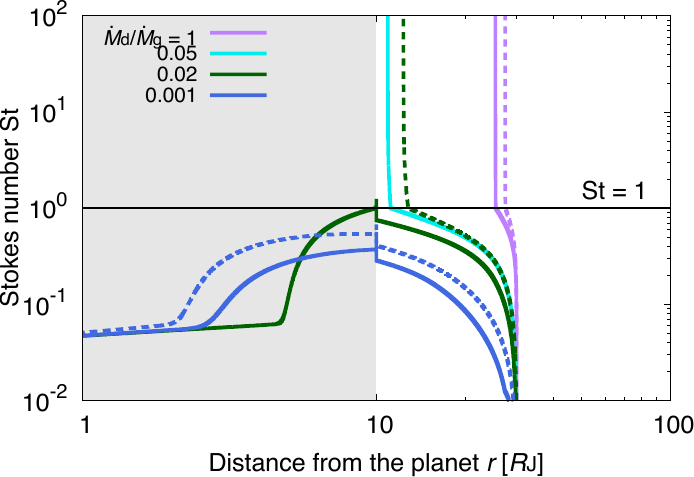}
\caption{Same as the lower left panel of Fig.~\ref{fig:evolution}, but with the non-uniform $\alpha_{\rm acc}$ described by Eq. (\ref{alpha_non-uni}) (solid curves). The dotted curves represent the profiles of the fiducial uniform $\alpha_{\rm acc}$ case ($\alpha_{\rm acc}=3.1\times10^{-3}$). The cyan curve represents the profile with the new critical value, $\dot{M}_{\rm d}/\dot{M}_{\rm g}=0.05$.}
\label{fig:non-uni}
\end{figure}

Figure~\ref{fig:non-uni} represents the dust evolution in the new modelled CPD. The shapes of the radial profiles of the Stokes number are similar to those in the fiducial model. The growth of the particles is less efficient than that of the fiducial case, and the critical value of the dust-to-gas ratio in the gas inflow, $\dot{M}_{\rm d}/\dot{M}_{\rm g}=0.05$, is also stricter. This is because $\eta$ is larger than that of the fiducial case (the bottom panel of Fig.~\ref{fig:alpha-Sigma-eta_non-uni}), which makes the drift speed of dust, $v_{\rm r}$, faster (Eq. (\ref{vr})), and then the growth efficiency is smaller (Eq. (\ref{growth})). However, this critical value is still $20$ times lower than that in turbulent CPDs \citep{shi17}, and the magnetic wind-driven accretion CPDs modelled in this section is still more suitable for the satellitesimal formation than turbulent CPDs.
 
\subsection{Possibility of the satellitesimal formation at the boundary of the wind-driven and gas inflow regions} \label{boundary}
\begin{figure}
\centering
\includegraphics[width=\columnwidth]{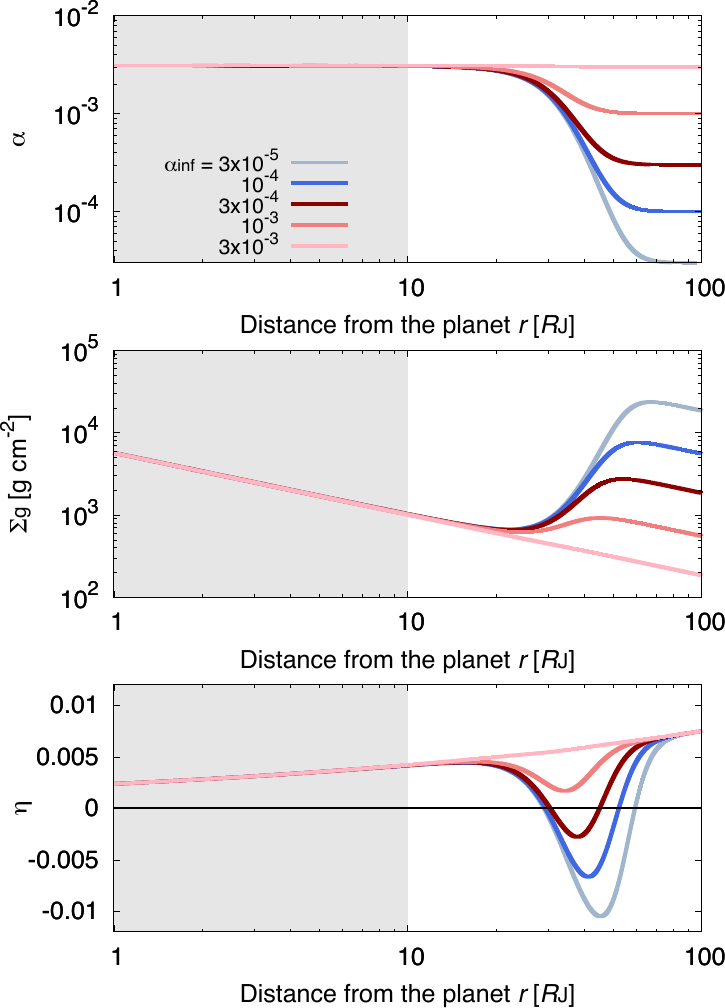}
\caption{Same as Fig.~\ref{fig:alpha-Sigma-eta_non-uni}, but with the non-uniform $\alpha_{\rm acc}$ described by Eq. (\ref{alpha_non-uni}) (solid curves). The range of the vertical axis of the top panel is also different. The colour variation shows that of the efficiency of the angular momentum in the gas inflow region, $\alpha_{\rm inf}$.}
\label{fig:alpha-Sigma-eta}
\end{figure}

We briefly investigate a possibility of the satellitesimal formation at the gas pressure bump formed by the transition of the inner wind-driven region ($r\leq30~R_{\rm J}$) and the outer gas inflow region ($30~R_{\rm J}< r$). We introduce a simple model being able to treat both of the regions. We assume $\alpha_{\rm acc}$ as
\begin{equation}
\alpha_{\rm acc}=\dfrac{\alpha_{\rm wind} - \alpha_{\rm inf}}{2}\left\{1-\tanh\left(\dfrac{r-r_{\rm tr}}{\Delta r_{\rm tr}}\right)\right\} + \alpha_{\rm inf},
\label{alpha_outer}
\end{equation}
where $\alpha_{\rm wind}$ and $\alpha_{\rm inf}$ are the efficiency of the angular momentum transport in the wind-driven and gas inflow regions, respectively. Considering that MRI does not work well in CPDs \citep[e.g.,][]{fuj14}, $\alpha_{\rm acc}$ could be lower in the gas inflow region than that of the wind-driven region\footnote{We do not consider the case that $\alpha_{\rm wind}<\alpha_{\rm inf}$, where the gas pressure bump is not formed, although it could be happen due to the stirring of the mid-plane gas by the inflow.}. Thus, we set $\alpha_{\rm wind}=3.1\times10^{-3}$ as a fixed value and change the value of $\alpha_{\rm inf}$ from $3\times10^{-5}$ to $3\times10^{-3}$ as a parameter. We also assume the boundary of the wind-driven and gas inflow regions as $r_{\rm tr}=30~R_{\rm J}$ and the width of the transition region as $\Delta r_{\rm tr}=10~R_{\rm J}$. We set the dust inflow orbital position as $100~R_{\rm J}$, and we consider that the gas disc of the whole dust evolving region (i.e., $r\leq100~R_{\rm J}$) is an accretion disc described by Eq. (\ref{sigmag}). The other equations of the model are the same with those used in Section \ref{formation}. Figure~\ref{fig:alpha-Sigma-eta} represents the radial distribution of $\alpha_{\rm acc}$ described by Eq. (\ref{alpha_outer}), and those of $\Sigma_{\rm g}$ and $\eta$ with it.

\begin{figure}
\centering
\includegraphics[width=\columnwidth]{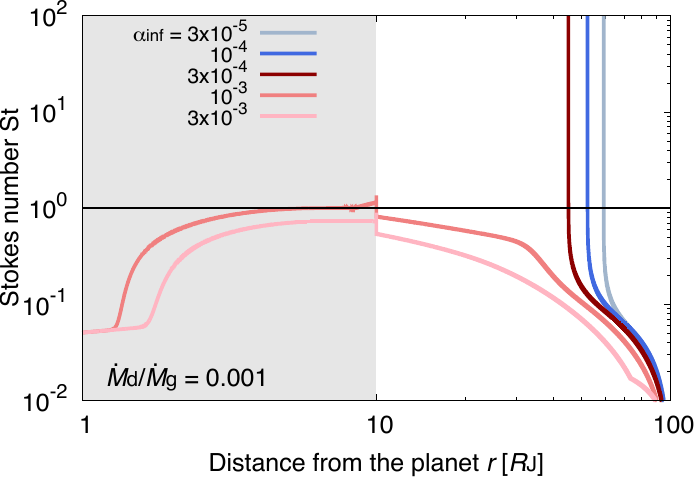}
\caption{Same as the lower left panel of Fig.~\ref{fig:evolution} but with the radial profile of $\alpha_{\rm acc}$ described by Eq. (\ref{alpha_outer}). The colour variation shows the various $\alpha_{\rm inf}$ taking the value from $3\times10^{-5}$ to $3\times10^{-5}$ (consistent with the colour variation of Fig.~\ref{fig:evolution}) with $\dot{M}_{\rm d}/\dot{M}_{\rm g}=0.001$.}
\label{fig:St-alphainf}
\end{figure}

Figure~\ref{fig:St-alphainf} represents the dust evolution with $\dot{M}_{\rm d}/\dot{M}_{\rm g} =0.001$. The profiles can be categorized into three types by the value of $\alpha_{\rm inf}$. First, when $\alpha_{\rm inf}\le3\times10^{-4}$, the dust particles can grow to satellitesimals at the gas pressure bump. We found that this formation condition is that the ratio of the pressure gradient force to the gravity of the planet, $\eta$, reaches zero. The bottom panel of Fig.~\ref{fig:alpha-Sigma-eta} shows $\eta$ is lower than zero when $\alpha_{\rm inf}\le3\times10^{-4}$. This is because, as $\eta$ goes to zero, the radial velocity of dust $v_{\rm r}$ goes to zero, which makes $\Sigma_{\rm d}$ larger toward infinity (see Eq. (\ref{continuous})). This increase of $\Sigma_{\rm d}$ causes the growth of dust to satellitesimals (see Eq. (\ref{growth})). The orbital position of the satellitesimal formation with each $\alpha_{\rm inf}$ is also consistent with the position where $\eta$ is zero. Second, when $\alpha_{\rm inf}=10^{-3}$ (the coral curves), the particles are piled up mildly at the bump, and the growth of particles is enhanced there. However, they do not grow to satellitesimals, because $\eta$ does not reach zero but only makes a small dent in its radial profile (see the bottom panel of Fig.~{\ref{fig:alpha-Sigma-eta}).} Third, when $\alpha_{\rm inf}=10^{-3}$ (the thin pink curves), the dust particles drift inward and do not grow to satellitesimals. In this case, $\alpha_{\rm acc}$ is uniform and the radial profile of $\eta$ is smooth. Thus, the shapes of the profiles of dust are same with those in the other sections except for the dust inflow position.

\section{Summary} \label{summary}
It has been considered that the satellitesimal formation via collisional growth of dust particles is difficult in turbulent CPDs because the particles drift toward the central planet before they grow large enough \citep{shi17}. The condition for satellitesimal formation against the drift barrier is that the dust-to-gas mass ratio of the inflow onto the CPDs is larger than unity ($\dot{M}_{\rm d}/\dot{M}_{\rm g}\geq1$), which is quite difficult to fulfill. However, if the disc is laminar, and the accretion of gas disc still works well, satellitesimals can form with lower dust-to-gas ratio. This is because the low gas density results in a slow drift velocity, and the weak vertical diffusion results in a high dust density on the mid-plane, leading to a high collision rate.

First, we carried out 3D local MHD simulations with all the three nonideal MHD effects. We found that the disc structure can be governed by the magnetic wind-driven accretion when the magnetic disc wind is launched from the surface of a CPD. The disc is laminar because the nonideal effects suppress MRI at the whole height of the disc.Furthermore, the magnetic pressure of the disc wind exceeds to the inflow Ram pressure in the region inside several tens $R_{\rm J}$. This suggests that the wind-accretion CPDs shown in this work can be stable with preventing the gas inflow from the PPD by the wind pressure. 

We then calculated the collisional growth and drift of the particles including the fragmentation effect. We set 1D disc models with the parameters which are consistent with the assumptions and results of the MHD simulations. As a result, the condition for the satellitesimal formation is $\dot{M}_{\rm d}/\dot{M}_{\rm g}\geq0.002$, which is 500 times smaller than the critical value in turbulent viscous accretion CPDs. When the KH instability is  considered, the condition becomes severer but is still $\dot{M}_{\rm d}/\dot{M}_{\rm g}\geq0.02$. Recent works show that the critical values of the dust-to-gas mass ratio of the inflow can be reached if enough amount of dust exists at the root of the gas inflow \citep{hom20,szu21}. For example, dust particles piled up at the gas pressure bump created by the planets make a lot of fragments by their mutual collision \citep{dra19}. We also show the possibility of the satellitesimal formation at the boundary of the inner wind-driven and outer gas inflow regions. The dust particles can be piled up and form satellitesimals at the gas pressure bump formed by the transition of $\alpha_{\rm acc}$ when it is steep enough. Therefore, the in-situ satellitesimal formation is plausible in laminar CPDs with magnetic-driven accretion.

\section*{Acknowledgements}
We thank the anonymous reviewer for the very constructive comments. We thank Xuening Bai for providing us with an improved version of the \texttt{Athena} code. We also thank Satoshi Okuzumi and Ryuki Hyodo for the valuable discussion. Parts of this work have been carried out within the framework of the National Centre of Competence in Research PlanetS supported by the Swiss National Science Foundation under grants 51NF40\_182901 and 51NF40\_205606. Y.S. also acknowledges the financial support of the Swiss National Science Foundation under grant 200020\_172746. This work was supported by JSPS KAKENHI Grant Numbers JP16J09590, JP17J10129, JP18H05222, JP21J00086, and JP22K14081. The MHD simulations were carried out on Cray XC50 at Center for Computational Astrophysics, National Astronomical Observatory of Japan.

\section*{Data availability}
The data underlying this article will be shared on reasonable request to the corresponding author. \texttt{Athena} is available at \url{https://princetonuniversity.github.io/Athena-Cversion/}.

\bibliographystyle{mnras}
\bibliography{text_accepted2}



\bsp	
\label{lastpage}
\end{document}